\documentclass{iopart}
\usepackage{graphicx}
\usepackage{amssymb}
\usepackage{color}
\usepackage{psfrag}
\usepackage{epsfig}
\usepackage{bbm}
\usepackage{bm}
\usepackage{dsfont}
\newcommand{\ket}[1]{| #1 \rangle}
\newcommand{\bra}[1]{\langle #1 |}
\newcommand{\rb}[1]{\left[ #1 \right]}

\newcommand{\beq}{\begin{eqnarray}}
\newcommand{\eeq}{\end{eqnarray}}

\newcommand{\eww}[1]{\langle #1\rangle}

\begin{document}
\title[Finite-frequency counting statistics of electron transport: Markovian Theory]{Finite-frequency counting statistics of electron transport: Markovian Theory}
\author{D. Marcos$^1$, C. Emary$^2$, T. Brandes$^2$ and R. Aguado$^1$}
\address{1-Departamento de Teor\'ia y Simulaci\'on de Materiales,
  Instituto de Ciencia de Materiales de Madrid, CSIC,
  Cantoblanco 28049, Madrid, Spain.}
 \address{2- Institut f\"ur Theoretische Physik,
  Hardenbergstr. 36,
  TU Berlin,
  D-10623 Berlin,
  Germany.}
\ead{david.marcos@icmm.csic.es}
\date{\today}

\begin{abstract}
We present a theory of frequency-dependent counting statistics of electron transport through nanostructures within the framework of Markovian quantum master equations. Our method allows the calculation of finite-frequency current cumulants of arbitrary order, as we explicitly show for the second- and third-order cumulants.  Our formulae generalize previous zero-frequency expressions in the literature and  can be viewed as an extension of MacDonald's formula beyond shot noise. When combined with an appropriate treatment of tunneling, using, e.~g. Liouvillian perturbation theory in Laplace space, our method can deal with arbitrary bias voltages and frequencies, as we illustrate with the paradigmatic example of transport through a single resonant level model. We discuss various interesting limits, including the recovery of the fluctuation-dissipation theorem near linear response, as well as some drawbacks inherent of the Markovian description arising from the neglect of quantum fluctuations.
\end{abstract}

\pacs{73.23.Hk, 72.70.+m, 02.50.-r, 03.65.Yz} \maketitle

\section{Introduction.}

Transport of electrons through nanoscopic conductors is a very powerful tool to learn about interactions and to characterize quantum systems \cite{Kouw97}. Examples include the quantum Hall effect \cite{Klitzing80}, weak localization \cite{Anderson58}, or universal conductance fluctuations \cite{AltshulerLeeStone85}. Transport processes are governed by tunneling events, which are stochastic in nature.  It is therefore natural to expect that the statistics of these tunneling events will be strongly influenced by interactions and quantum effects. Interestingly, these statistics, which can be analyzed by studying current fluctuations, contain a great deal of new information beyond that provided by dc transport \cite{Bla00, NazarovBook}. In particular, the second-order current correlation function (noise), can be used to determine the effective charge \cite{Kane94dePicciotto97Glattli97} and the statistics of the quasiparticles \cite{Henny-Oliver99}, and to reveal information on the transmission properties of the conductor \cite{Bla00}. Moreover, current correlations can be used to learn about entanglement \cite{Beenakker06,Klich-Levitov09}, quantum coherence \cite{Kiesslich07}, and the deep connection that exists with fluctuation theorems \cite{Exposito1,Utsumi,Buttiker1}. 
Further information can be gained from noise at high frequencies which is valuable for extracting internal energy scales in systems such as quantum dots \cite{Aguado04}, spin-valves \cite{Braun-etal}, Cooper Pair Boxes \cite{Choi}, diffusive wires \cite{Nag98} or chaotic cavities \cite{Nag04}. 

While most of the work on noise is theoretical (in particular at high frequencies), the field of noise and counting statistics is producing a great deal of experimental breakthroughs, including measurements of high order cumulants \cite{Reu03,Rez05,Lu03,Gus06,Fuj06,experiments3,experiments4,experiments5,experiments6,experiments7}. Owing to this experimental progress, noise measurements at high frequencies, which until recently were scarce, are now possible  \cite{freq-exp1,freq-exp2,freq-exp3,freq-exp4,freq-exp5,freq-exp6,freq-exp7,freq-exp8,freq-exp9}.

A  proper treatment of fluctuations in non-equilibrium transport is needed to address the problems listed above.  While the list of theoretical available is too large to be given here, it is safe to say that they can be divided roughly into three families: the scattering approach \cite{Butt95,Bla00}, the Keldysh Green's functions method \cite{Meir92Jauho}, and the various quantum master equation (QME) treatments \cite{Cohen} (for a recent overview of transport in this last context see e.g. Ref~\cite{Timm}). 
A theory of counting statistics of electron transport was first formulated by Levitov and Lesovik for noninteracting electrons using the scattering formalism \cite{Lev}, and later
works enabled the treatment interacting problems \cite{Nazarov99,Bel01}. QME
approaches followed \cite{Bag03}, and prooved particularly useful for studying
systems in the Coulomb Blockade regime.  Recent avances within this last scheme also involve studies of the counting statistics including non-Markovian dynamics \cite{BraggioNM,FlindtNM1,FlindtNM2,Marcos10}.  

In Ref.~\cite{Emary07} we developed a method for calculating high-order current correlations at finite frequencies in the context of Markovian QMEs, and the aim of this paper is to significantly extend that work. We provide a detailed derivation of our multi-time generating function, Eq. (\ref{CGF}). We present a new approach to derive finite-frequency cumulants from this expression, noticing that only the stationary solution of the problem is required.  
We give analytical expressions for the second and third-order current cumulants (noise and skewness respectively). These results generalize previous zero-frequency expressions in the literature and recover the finite frequency shot noise expressions \cite{FlindtPhysE,Lambert07} obtained using the MacDonald's formula \cite{MacDonnald48}. Our method can thus be viewed as a generalization of this formula, as it allows us to obtain high-order spectra such as the frequency-dependent skewness (Eq. (\ref{skewnessformula})). To illustrate the formalism we study the case of a single resonant level (SRL) model, and compare it with the exact solution and with the non-equilibrium version of the fluctuation-dissipation theorem, derived in various approaches, such as for tunnel junctions, or for the weak cotunneling regime in quantum dots \cite{Lev04,Dahm69,rogovin-scalapino,Suk01}.

The paper is organized as follows. In section \ref{TheorySec}, we present our formalism of finite-frequency cumulants in the context of Markovian QMEs. Subsections \ref{QMEsec} and \ref{FCSsec} are devoted to establish the general framework of full counting statistics. In subsection \ref{FCSwsec} we derive a multitime cumulant generating function. Subsection \ref{S2S3sec} shows how to obtain finite-frequency cumulants of the current distribution. Here, exact equations for the frequency-dependent noise and skewness are given. We end this part with special emphasis on how to calculate the counting statistics of the ``total" and ``accumulated'' currents (subsection \ref{FCStotaccum}). We explicitly show how both current correlations and charge correlations can be calculated.
In section \ref{Resultssec}, we study the example of a SRL model, providing spectra for the frequency-dependent noise and skewness and a detailed comparison with the fluctuation-dissipation theorem, the finite-frequency version of the non-equilibrium fluctuation-dissipation theorem, and the exact solution of the SRL model. First we focus on the zero-frequency case (section \ref{FCSresults}), where the general behaviour of noise and skewness is presented as a function of different system parameters. In section \ref{FCSwresults} we extend this study to the finite-frequency case. Interestingly, we show that, even though the theory does not contain quantum fluctuations, the Markovian limit is basically exact in transport configurations, level within the bias voltage window, as long as $\hbar\omega \gg eV$ or $\hbar\omega\ll eV$. In intermediate situations, where $\hbar\omega \sim eV$, or with the level outside the bias windows, the Markovian limit fails at finite frequencies due to its lack of quantum fluctuations. We also demonstrate that the noise spectra for particle currents and the ones for total currents significantly deviate from each other, even for large asymmetric coupling to the leads, namely  
$\Gamma_R/\Gamma_L \neq 1$.
In section \ref{ConclusionsSec} we summarize our results. Most of the technical details and intermediate steps of the derivations in section \ref{TheorySec} are discussed in detail in \ref{appDiagrams}, where we also present a diagrammatic technique to arrive to the expressions for the cumulants shown in sec. \ref{S2S3sec}. In section \ref{appKernel} we describe how to calculate the kernel of the QME to lowest (sequential) order using perturbation theory in the Liouville space.

\section{Theory} \label{TheorySec}

\subsection{Quantum master equation} \label{QMEsec}

We are interested in phenomena which can be described by an equation of the type
\beq \label{LiouvilleEq}
\dot{\rho}(t) = \mathcal{L} \rho(t),
\eeq
where $\mathcal{L}$ is the so called Liouvillian, governing the evolution, and $\rho$ is the density operator of the total system. Specifically, our theory will be useful to processes amenable to the counting of a classical stochastic variable $n$, which can be, for example, the number of particles that have undergone a particular process in the system.
We will focus in particular in transport systems, consisting on a central
region, with a known set of many-body eigenstates $\{\ket{a}\}$, and respective eigenenergies $\{E_a\}$, attached to non-interacting electronic leads at different chemical potentials.
This set-up can be described by a Hamiltonian that takes the form
$\mathcal{H}=\mathcal{H}_S + \mathcal{H}_R + \mathcal{H}_T$, where
$\mathcal{H}_S$ and $\mathcal{H}_R$ refer to system and leads
respectively, and $\mathcal{H}_T$ is the coupling between them. The different terms can be written as
\begin{eqnarray}
\mathcal{H}_S &=& \sum_a E_a \ket{a}\bra{a} \label{Hs}\\
\mathcal{H}_R &=& \sum_{\eta,\alpha} \left( \varepsilon_{\eta\alpha}
+ \mu_{\alpha}
\right) c_{\eta\alpha}^{\dagger} c_{\eta\alpha} \label{Hr}\\
\mathcal{H}_T &=& \sum_{\eta,\alpha,m} {\cal V}_{\eta\alpha m}
c_{\eta\alpha}^{\dagger} d_m + \mathrm{H.c.} \label{Hv}
\end{eqnarray}
Here $c_{\eta\alpha}^{\dagger}$ creates an electron
with quantum numbers $\eta$ in lead $\alpha$, and $d_m$ annihilates
an electron from site $m$ in the central region.  $\varepsilon_{\eta\alpha}$ are the eigenenergies of the electrons in the lead $\alpha$, ${\cal V}_{\eta\alpha m}$ is the coupling energy between a state in contact $\alpha$ and the level $m$ in the system. $\mu_{\alpha}$ the chemical potential of lead $\alpha$, and that allows the system to be driven out of equilibrium. 

Given the Hamiltonian, the full evolution can be written in terms of the Liouvillian by using the von Neumann
equation \beq \label{vonNeumann} \dot{\rho}(t) = -\frac{i}{\hbar} \left[
\mathcal{H}, \rho(t) \right] \equiv \mathcal{L} \rho(t). \eeq
We are actually interested in the dynamics concerning the central system. We therefore trace out the reservoir degrees of freedom, arriving at an equation of motion for the reduced density operator $\rho_\mathrm{S}(t)\equiv\mathrm{Tr_{R}}\{\rho(t)\}$, that, if we assume the dynamics of the reservoirs to be much faster than that of the central system, can be approximated as a Markovian master equation
\beq \label{EOMt}
\dot{\rho}_S(t) =  \mathcal{W}\rho_S(t), 
\eeq
where ${\cal W}$ is a kernel driving the dynamics of the system. The charge flow through the conductor is governed by the stochastic hopping of electrons in and out of the central region. This processes are susceptible to classical counting and thus the reduced density operator can be unravelled into components $\rho_\mathrm{S}(n_{\alpha},t)$, corresponding to having $n_{\alpha}=...,-1,0,1,...$ extra electrons in lead $\alpha$ \cite{Cook81,Plenio98}. The kernel can also be split as ${\cal W}={\cal W}_0+\sum_{\pm}{\cal W}_{\pm}^{\alpha}$, where ${\cal W}_{\pm}^{\alpha}$ refers to the physical process in which one electron in created (+) or annihilated (-) at lead $\alpha$, and ${\cal W}_0$ corresponds to the part in which no tunneling processes take place in that lead. It can be shown that $\rho_\mathrm{S}(n_{\alpha},t)$ fulfills the equation (see e.g. \cite{flindtepl}):
\beq \label{nQME}
\dot{\rho}_\mathrm{S}(n_{\alpha},t) = {\cal W}_0 \rho_\mathrm{S}(n_{\alpha},t) + {\cal W}_{+}^{\alpha} \rho_\mathrm{S}(n_{\alpha} - 1,t) + {\cal W}_{-}^{\alpha} \rho_\mathrm{S}(n_{\alpha} + 1,t);
\eeq
valid provided that only single particle tunneling processes occur. Although Eq. (\ref{nQME}) focuses on the single counting at a particular lead $\alpha$, it can be generalized to account for tunneling processes of $k$ particles at the different system-reservoir junctions:
\beq \label{nQMEgeneralized}
\dot{\rho}_\mathrm{S}(n_1,\ldots,n_M,t) = && {\cal W}_0 \rho_\mathrm{S}(n_1,\ldots,n_M,t) \nonumber\\ &&+ \sum_{\alpha,k,\pm} {\cal W}_{\pm}^{\alpha,k}\rho_\mathrm{S}(n_1,\ldots,n_{\alpha}\mp k,\ldots,n_M,t)
\eeq
where ${\cal W}_0$ is the part in which the number of particles is not changed in the central region, $M$ the number of leads, and $k$ labels the process in which $k$ particles ``jump'' at a time.

Unfortunately, solving equation (\ref{nQMEgeneralized}) in the $n$-space requires truncation to a certain $n$ and diagonalization of a tridiagonal matrix. It is therefore more convenient to solve it taking its Fourier transform. Multiplying (\ref{nQMEgeneralized}) by $e^{in_1\chi_1}\ldots e^{in_M\chi_M}$ and summing over $n_1,\ldots,n_M$ we obtain
\beq \label{EOMchit} \dot{\rho}_S(\chi,t) =
\mathcal{W}(\chi)\rho_S(\chi,t), \eeq
Here the counting field $\chi$ refers implicitly to all counting fields, and we have defined $\rho_\mathrm{S}(\chi,t)\equiv\sum_{n_1,\ldots,n_M} e^{in_1\chi_1}\ldots e^{in_M\chi_M} \rho_\mathrm{S}(n_1,\ldots,n_M,t)$ and ${\cal W}(\chi)\equiv{\cal W}_0+\sum_{\alpha,\pm}{\cal W}_{\pm}^{\alpha}e^{\pm i\chi_{\alpha}}$.
For a time-independent kernel, the solution to Eq.~(\ref{EOMchit}) is 
\beq \label{rhochit}
{\rho}_\mathrm{S}(\chi,t)=\Omega(\chi,t-t_0){\rho}_\mathrm{S}(\chi,t_0),
\eeq
with the time evolution operator $\Omega(\chi,t-t_0):=e^{{\cal W}(\chi)(t-t_0)}$.

\subsection{Full Counting Statistics} \label{FCSsec}

Importantly, the knowledge of the system's density operator resolved in $n$ allows us to obtain the the full counting statistics (FCS) of the system, that is the probability distribution $P(n_1,\ldots,n_M,t)$ of the number of electrons
transmitted through the system-lead junctions. This is accomplished by noting that \beq\label{Pn} P(n_1,\ldots,n_M,t) = \mathrm{Tr} \{ \rho_S(n_1,\ldots,n_M,t) \}.\eeq 
Transforming the probability distribution to the $\chi$-space, we have the moment generating function (MGF)
\beq\label{MGF-CGF}{\cal
G}(\chi,t)=\sum_n P(n,t) e^{in\chi},\eeq
where now $n$ refers implicitly to $n_1,\ldots,n_M$. Using Eq.~(\ref{rhochit}) one has
\beq \label{MGF}
{\cal G}(\chi,t)=\mathrm{Tr} \left\lbrace \Omega(\chi,t-t_0){\rho}_\mathrm{S}(t_0) \right\rbrace,
\eeq
equation that was already derived by Bagrets and Nazarov \cite{Bag03}.
The $N$-th. derivative of the MGF with respect to $\chi$ gives the $N$-th moment of the distribution of the number of particles that have gone in or out a particular lead $\alpha$:
\beq
\langle n_{\alpha}^N (t) \rangle=\frac{\partial^N {\cal
G}(\chi,t)}{\partial(i\chi_{\alpha})^N}|_{\chi\rightarrow
0}.
\eeq
When equation (\ref{MGF}) is used, averages with respect to the stationary state are established by taking 
${\rho}_\mathrm{S}(t_0)={\rho}_\mathrm{S}^{stat}$ (defined by
${\mathcal W}{\rho}_\mathrm{S}^{stat}=0$). This means that counting will start at
a time $t_0$ in which the system has reached its steady state, and therefore the fluctuations we study are around this state
\footnote{In the following, all the averages
will be taken with respect to this steady state. An average in 
the Liouville space will be therefore written as $\langle A \rangle = \bm{t}_T\cdot A \cdot \bm{\rho}_\mathrm{S}^{stat} \equiv \langle\!\langle \tilde{0} | A | 0 \rangle\!\rangle$, where $| 0 \rangle\!\rangle \equiv \bm{\rho}_\mathrm{S}^{stat}$ is the normalized stationary system density matrix (written as a vector), and $\bm{t}_T\equiv \langle\!\langle \tilde{0} |$ is the transposed trace vector that sums over the population degrees of freedom.}.

The moments of the current distribution can be calculated as \footnote{Throughout the paper we will use $e$ (electron charge) $=$ $k$ (Boltzman's constant) $=$ $\hbar$ (Planck's constant/$2\pi$) = 1.}
\beq
\langle I_{\alpha}^N (t) \rangle = \frac{d}{dt} \langle n_{\alpha}^N (t) \rangle.
\eeq
This relation is important as it relates the stochastic variable $n$ with the current of particles flowing through the system. Even though the current studied here is a classical variable, it contains quantum effects present in the system. In the formalism, these are inherited from the Liouvillian operator in equation (\ref{LiouvilleEq}).
Generally, we are interested in the cumulants, rather than the moments, of the current distribution.  These can be obtained from the derivatives of the cumulant generating function (CGF), defined by ${\cal F}(\chi,t) := ln {\cal G}(\chi,t)$. Therefore we have
\beq
\langle
I_{\alpha}^N\rangle_c=\frac{d}{dt}\frac{\partial^N {\cal
F}(\chi,t)}{\partial(i\chi_{\alpha})^N}|_{\chi\rightarrow
0,t\rightarrow\infty},
\eeq
where $\langle \ldots \rangle_c$ denotes cumulant average \cite{Kubo62} and the limit $t\to\infty$ ensures that the average is performed in the stationary state. Also, notice that the probability distribution itself can be obtained by inverse Fourier transform of the MGF.

From the $\chi$-independent kernel of the reduced QME (\ref{EOMt}), the $\chi$-dependence leading to equation (\ref{EOMchit}) can be actually introduced in a simpler way than resolving the density operator in $n$ and taking the Fourier transform. As we describe in \ref{appKernel}, it is enough to include counting fields in the appropriate tunneling terms of the Kernel \cite{SETs}, and this procedure is fully equivalent to solving a generalized Von Neumann equation: \beq \label{EOMchi} \dot{\rho}(\chi,t) = -\frac{i}{\hbar}
(\mathcal{H}^+(t)\rho(\chi,t)-\rho(\chi,t)\mathcal{H}^-(t)), \eeq in
which the time evolution in the forward (+) and backward (-) Keldysh
part of the real time axis is governed by \emph{different}
Hamiltonians \cite{Lev04}, specifically $\mathcal{H}^{\pm}_T =
\sum_{\eta,\alpha,m} {\cal V}_{\eta\alpha m} e^{\pm
i\chi/2}c_{\eta\alpha}^{\dagger} d_m + \mathrm{H.c.}$. Tracing out the 
reservoir degrees of freedom in equation (\ref{EOMchi}), one can get 
equation (\ref{EOMchit}) and proceed to obtain the FCS
of the system.

\subsection{Finite-frequency full counting statistics} \label{FCSwsec}

In this paper we want to study correlations at finite
frequencies, for which the scheme presented above has to be
generalized. To this end, one needs to consider a \emph{joint probability
distribution}, $P(n_1,t_1;\ldots;n_N,t_N)$ defined as the
probability that $n_1$ electrons have undergone a particular process
 after a time $t_1$, $n_2$ electrons after a time $t_2$, etc. Here we focus
for simplicity on a particular lead, and denote $n$ as the number of particles
transferred to (from) it, with associated counting field $\chi$ ($-\chi$). Being straightforward to include processes at different leads.

The connection between this joint probability and the density 
operator (analog to Eq.~(\ref{Pn})) is not straightforward. To connect them
we first need to specify a prescription for the symmetrization of the cumulants and the probability distribution. 
This prescription actually depends on the detection scheme \cite{DetecSym1,QPC3,DetecSym2}. Here we assume that ``classical'' detection
is carried out, so the detector is incapable of distinguishing emission from absorption. This means that the results we will present correspond to the fully-symmetrized version of the power spectrum
\beq \label{currentcorrfreq}
S^{(N)}(\omega_1,\ldots,\omega_N):= \int_{-\infty}^\infty dt_1\ldots dt_N && e^{-i\omega_1 t_1}\ldots e^{-i\omega_Nt_N} \nonumber\\ &&\times {\cal T}_S \langle I(t_1) \ldots
I(t_N) \rangle_c,
\eeq
where ${\cal T}_S$ is the symmetrization operator, that sums over all possible time (or frequency) switchings, that is, we have for example ${\cal T}_S \langle I(t_1)I(t_2) \rangle = \langle I(t_1)I(t_2) \rangle + \langle I(t_2)I(t_1) \rangle$.

The spectrum (\ref{currentcorrfreq}) can be derived from a $N$-time (symmetrized) CGF ${\cal F}^{(N)}$, defined by
\beq e^{\mathcal{F}^{(N)}[\bm{\chi},\bm{t}]} = {\cal G}^{(N)} [\bm{\chi},\bm{t}] :=
\sum_{n_1,\ldots,n_N} e^{in_1\chi_1+\ldots+in_N\chi_N}
P^{(N)}[\bm{n},\bm{t}], \nonumber \eeq where
$\bm{\chi}:=(\chi_1,\ldots,\chi_N)_T$, $\bm{t}:=(t_1,\ldots,t_N)_T$, $\bm{n}:=(n_1,\ldots,n_N)_T$, $P^{(N)}$ refers to the symmetrized joint probability, ${\cal G}^{(N)}$ to the multitime MGF, and the subscript $T$ to the transpose of a column vector. That is, we have
\beq
S^{(N)}(\omega_1,\ldots,\omega_N)= && \int_{-\infty}^\infty dt_1\ldots dt_N e^{-i\omega_1 t_1}\ldots e^{-i\omega_Nt_N} \nonumber\\ && \times
\partial_{t_1} \ldots \partial_{t_N}
\partial_{i\chi_1} \ldots \partial_{i\chi_N} \mathcal{F}^{(N)}[\bm{\chi},\bm{t}] \;
\big\vert_{\bm{\chi}=\bm{0}} .\label{currentcorrtime} \eeq
And using the property of the Fourier transform of a derivative we get
\beq
S^{(N)}(\omega_1,\ldots,\omega_N)= && (i\omega_1) \ldots (i\omega_N) \int_{-\infty}^\infty dt_1\ldots dt_N e^{-i\omega_1 t_1}\ldots e^{-i\omega_Nt_N} \nonumber\\ && \times
 \partial_{i\chi_1} \ldots \partial_{i\chi_N} \mathcal{F}^{(N)}[\bm{\chi},\bm{t}] \;\big\vert_{\bm{\chi}=\bm{0}} .
\eeq
Both the probability $P^{(N)}[\bm{n},\bm{t}]$ and the CGF $\mathcal{F}^{(N)}[\bm{\chi},\bm{t}]$ can be calculated from the density operator and the Kernel ${\cal W}$ if we use the Markovian approximation in the coupling with central system--reservoir coupling. Within that limit we have the evolution local in time given by (\ref{rhochit}) and also the factorization property
\beq \label{Pgreater}
P^>(n_1,t_1;\ldots;n_N,t_N) = && P(n_1,t_1) P(n_2,t_2 | n_1,t_1) \nonumber\\ && \times \ldots P(n_N,t_N | n_{N-1},t_{N-1}),
\eeq
where the symbol $>$ constraints the times to $t_k>t_{k-1}$. Notice that as we are considering the totally symmetric correlation function, we need to consider $P^{(N)}[\bm{n},\bm{t}] = {\cal T} P^>(n_1,t_1;\ldots;n_N,t_N)$, where ${\cal T}$ is the time-ordering operator \cite{Emary07}. $P(n,t | n',t')$ is the conditional probability of counting $n$ electrons after time $t$ provided that we counted $n'$ electrons after time $t'$, and can be computed as
\beq \label{Pcond}
P(n,t | n',t') = \frac{\mathrm{Tr}\lbrace \Omega(n-n',t-t')\rho_\mathrm{S}(n',t')\rbrace}{\mathrm{Tr}\lbrace \rho_\mathrm{S}(n',t') \rbrace}
\eeq
where the normalization in the denominator accounts for the collapse of the state due to the measurement, as given by the von Neumann's projection postulate \cite{Korotkov01}. $\Omega(n,t)$ is the propagator in the $n$-space, that is,
\beq\label{evolrho}{\rho}_\mathrm{S}(n,t)=\sum_{n'}\Omega(n-n',t-t'){\rho}_\mathrm{S}(n',t'),\eeq and can be extracted from equation (\ref{nQMEgeneralized}) or by inverse Fourier transform of the propagator in the $\chi$-space:
\beq
\Omega(n,t) = \int \frac{d\chi}{2\pi} e^{-in\chi} \Omega(\chi,t)
\eeq
An expression for the joint probability distribution in terms of propagators can be then derived using (\ref{Pgreater}) together with (\ref{Pcond}) and (\ref{evolrho}). Alternatively, it can be obtained using the Chapman-Kolmogorov property for Markovian evolutions, from which we have
\beq
P(n_N;t_N) && = \mathrm{Tr} \sum_{n_1,\ldots,n_{N-1}} \Omega(n_N-n_{N-1};t_N-t_{N-1}) \nonumber\\ &&\times \Omega(n_{N-1}-n_{N-2};t_{N-1}-t_{N-2}) \ldots \Omega(n_1;t_1) \rho_\mathrm{S}(t_0)
\eeq
As we also have $P(n_N;t_N)=\sum_{n_1,\ldots,n_{N-1}} P^>(n_1,t_1;\ldots;n_N,t_N)$, reminding that ${\rho}_\mathrm{S}(t_0)={\rho}_\mathrm{S}^{stat}$, we find
\beq \label{jointPn}
P^{(N)}[\bm{n},\bm{t}] = \mathcal{T}\Big\langle
  \prod_{k=1}^N
\Omega(\nu_{N-k},\tau_{N-k})
  \Big\rangle,
\eeq
where ${\nu}_{k}:=n_{k+1}-n_k$, ${\tau}_{k}:=t_{k+1}-t_k$ and $\langle \bullet \rangle := \mathrm{Tr}\{\bullet {\rho}_\mathrm{S}^{stat} \}$. Transforming expression (\ref{jointPn}) to the $\chi$-space, we find the CGF
\beq\label{CGF}
 {\mathcal{F}^{(N)}[\bm{\chi},\bm{t}]}
 =
 \ln \mathcal{T}\Big\langle
  \prod_{k=1}^N
\Omega(\tilde{\chi}_k,\tau_{N-k})
  \Big\rangle,
  \eeq
being $\tilde{\chi}_k:=\sum_{i=N+1-k}^N \chi_i$. The structure in Eq. (\ref{CGF}) is
encountered in many branches of physics such as statistical physics
and field theory, where \emph{connected} correlation functions are
obtained by taking derivatives of the logarithm of the corresponding
generating functional (the partition function, the $S$-matrix, etc). Note
in particular the analogy with the partition function presented in Ref.~\cite{Kubo62}.

\subsection{Finite-frequency cumulants} \label{S2S3sec}

Equation (\ref{CGF}) permits us to obtain frequency-dependent current cumulants to arbitrary order. This was used in Ref.~\cite{Emary07} to study the second and third cumulant in single and double quantum dot models. Explicit derivatives of (\ref{CGF}) and the eigen-decomposition of the Kernel was used then to that end.
In this subsection we show that only the stationary solution of the problem (solution to an algebraic equation) is needed to compute the finite-frequency current cumulants. We give analytical expressions (valid within the Markovian approximation) for the noise (second cumulant) and skewness (third cumulant) of the distribution of charge flowing through a conductor.

Let us decompose the Fourier transform in equation (\ref{currentcorrfreq}) into a set of Laplace transforms (defined as $f(z):=\int_0^{\infty} dt e^{-zt} f(t)$), and the cumulant averages in terms of moments (c.f. for example Eq.~(2.8) in Ref~\cite{Kubo62}). Doing this we find \footnote{Notice that eqs. (\ref{LaplaceS1})-(\ref{LaplaceS3}) can also be derived if the derivatives of the CGF are decomposed in terms of derivatives of MGFs. For example, for $N=3$ we have ${\mathcal{F}^{(3)}_{123}}={\mathcal{G}^{(3)}_{123}}-{\mathcal{G}^{(3)}_{1}}{\mathcal{G}^{(3)}_{23}}-{\mathcal{G}^{(3)}_{2}}{\mathcal{G}^{(3)}_{13}}-{\mathcal{G}^{(3)}_{3}}{\mathcal{G}^{(3)}_{12}}+2{\mathcal{G}^{(3)}_{1}}{\mathcal{G}^{(3)}_{2}}{\mathcal{G}^{(3)}_{3}}$, with $f_{i}:=\partial_{\chi_i} f\vert_{\chi_i=0}$, $f_{ij}:=\partial_{\chi_i}\partial_{\chi_j} f\vert_{\chi_i,\chi_j=0}$, $f_{ijk}:=\partial_{\chi_i}\partial_{\chi_j}\partial_{\chi_k} f\vert_{\chi_i,\chi_j,\chi_k=0}$.}
\beq \label{LaplaceS1}
S^{(1)>}(z_1) = S_m^{(1)>}(z_1) \;,
\eeq
\beq \label{LaplaceS2}
S^{(2)>}(z_1,z_2) = S_m^{(2)>}(z_1,z_2) - \left( \frac{-1}{z_1} \right) \left( \frac{-1}{z_2} \right) \langle I \rangle^2,
\eeq
\beq \label{LaplaceS3}
S^{(3)>}(z_1,z_2,z_3)
&=& S_m^{(3)>}(z_1,z_2,z_3) \nonumber\\ &&- \left( \frac{-1}{z_1} \right) \langle I \rangle S_m^{(2)>} (z_2,z_3) \nonumber\\ &&- \left( \frac{-1}{z_2} \right) \langle I \rangle S_m^{(2)>} (z_1,z_3) \nonumber\\ &&- \left( \frac{-1}{z_3} \right) \langle I \rangle S_m^{(2)>} (z_1,z_2) \nonumber\\ &&+ 2 \left( \frac{-1}{z_1} \right)  \left( \frac{-1}{z_2} \right)  \left( \frac{-1}{z_3} \right) \langle I \rangle^3.
\eeq
The notation ``$>$'' denotes the unsymmetrized correlation function corresponding to the time ordering $t_N>\ldots>t_2>t_1$. Symmetrization in the frequency space implies adding the part corresponding to negative $z$ and summing over all the possible switchings of frequencies.
The subscript $m$ means moment. These can be obtained as
\beq
S_m^{(N)>}(z_1,\ldots,z_N) = z_1 \ldots z_N
\partial_{i\chi_1} \ldots \partial_{i\chi_N} \mathcal{G}^{(N)>}[\bm{\chi},\bm{z}] \;
\big\vert_{\bm{\chi}=\bm{0}}, \label{currentcorr}
\eeq 
with $\bm{z}\equiv(z_1,\ldots,z_N)_T$ and \footnote{In the frequency domain, the prescription $>$ can be taken similarly, that is $z_N>\ldots>z_2>z_1$, and finally symmetrize the result.} \beq \label{GN}\mathcal{G}^{(N)>} [\bm{\chi},\bm{z}] = \Big\langle \prod_{k=1}^N
\Omega(\tilde{\chi}_k,\tilde{z}_{k}) \Big\rangle^>,\eeq and $\Omega(\chi,z)\equiv [z-{\cal W}(\chi)]^{-1}$.

The advantage of having moment averages is that we can use a diagrammatic technique (see \ref{appDiagrams}) to easily obtain the correlation functions.
Symmetrizing $S^{(N)>}[\bm{z}]$ and evaluating it at $\bm{z}=i\bm{\omega}$ (being $\bm{\omega}\equiv (\omega_1,\ldots,\omega_N)_T$) we find that $S^{(N)}[\bm{\omega}]$ is proportional to $\delta(\omega_1+\ldots+\omega_N)$, as required by time-translational invariance. Defining the jump superoperators ${\cal
J}_{\chi}:=[{\cal W}(\chi)-{\cal W}(\chi=0)]$ and their derivatives $\mathcal{J}_0^{(n)}:=\partial_{\chi}^n\mathcal{J}_{\chi}|_{\chi=0}$, we arrive to the following expressions for the current, noise and skewness of the current distribution (c.f. \ref{appDiagrams} for details):
\beq
\label{currentformula} i I_{stat} = \langle \mathcal{J}_0^{(1)}
\rangle, \eeq
\beq
\label{noiseformula} i^2 S^{(2)} (\omega) = \langle {\cal
J}_{0}^{(2)} \rangle + \langle {\cal J}_0^{(1)} {\Omega}_0(i\omega)
{\cal J}_0^{(1)} \rangle + \langle {\cal J}_0^{(1)} {
\Omega}_0(-i\omega) {\cal J}_0^{(1)} \rangle, \eeq
\beq
\label{skewnessformula} i^3 S^{(3)} (\omega,\omega') &=&
\langle {\cal J}_{0}^{(3)} \rangle + \langle{\cal
J}_0^{(2)}{\Omega}_0(i\omega){\cal J}_0^{(1)} \rangle +
\langle{\cal J}_0^{(2)}{\Omega}_0(i\omega'-i\omega){\cal J}_0^{(1)}
\rangle \nonumber\\ &&+ \langle{\cal J}_0^{(2)}{\Omega}_0(-i\omega'){\cal J}_0^{(1)} \rangle +
 \langle{\cal J}_0^{(1)}{\Omega}_0(-i\omega){\cal J}_0^{(2)}
\rangle \nonumber\\ &&+ \langle{\cal J}_0^{(1)}{\Omega}_0(i\omega'){\cal J}_0^{(2)} \rangle + \langle{\cal J}_0^{(1)}{\Omega}_0(i\omega-i\omega'){\cal J}_0^{(2)} \rangle \nonumber \\ &&+
\langle {\cal J}_0^{(1)}{\Omega}_0(-i\omega){\cal J}_0^{(1)}{\Omega}_0(-i\omega'){\cal J}_0^{(1)}\rangle \nonumber \\ &&+ \langle {\cal
J}_0^{(1)}{\Omega}_0(i\omega'){\cal J}_0^{(1)}{\Omega}_0(i\omega){\cal J}_0^{(1)}\rangle \nonumber \\ &&+ \langle
{\cal J}_0^{(1)}{\Omega}_0(-i\omega){\cal J}_0^{(1)}{\Omega}_0(i\omega'-i\omega){\cal J}_0^{(1)}\rangle \nonumber \\ &&+
\langle {\cal J}_0^{(1)}{\Omega}_0(i\omega'){\cal J}_0^{(1)}{\Omega}_0(i\omega'-i\omega){\cal J}_0^{(1)}\rangle \nonumber \\ &&+
\langle {\cal J}_0^{(1)}{\Omega}_0(i\omega-i\omega'){\cal
J}_0^{(1)}{\Omega}_0(-i\omega'){\cal J}_0^{(1)}\rangle \nonumber\\ &&+
\langle {\cal J}_0^{(1)}{\Omega}_0(i\omega-i\omega'){\cal
J}_0^{(1)}{\Omega}_0(i\omega){\cal J}_0^{(1)}\rangle, \eeq being
$\Omega_0(z):=[z-{\cal W}(\chi=0)]^{-1}$.
These equations generalize the zero-frequency results found in Ref. \cite{flindtepl} (c.f. their eqs Eqs. (7) and (8)) to finite frequencies (see \ref{appDiagrams} for the zero-frequency limit of (\ref{currentformula})-(\ref{skewnessformula})).
Results for higher-order cumulants can be similarly obtained.

The relation between cumulants and moments can be formally expressed more generally at the level of the generating function. To do this one should follow the derivation by Kubo \cite{Kubo62}, making use of the property $\langle exp(\sum_i n_i\chi_i)\rangle=exp\{\langle exp(\sum_i n_i\chi_i)-1\rangle_c\}$ in our context, arriving to a similar result to (7.25) in Ref.~\cite{Kubo62}.
This allows for the calculation of frequency-dependent cumulants of the current distribution up to any order, reproducing in particular the results presented above. If a diagrammatic expansion in the Liouvillian space is used \cite{Schoeller09,Leijnse08,Emary09}, cumulant averages become particularly useful since one can then keep only \textit{connected} diagrams as those contributing to the average, in a similar way that this is done in quantum field theory.

\subsection{FCS of total and accumulated currents\label{total}} \label{FCStotaccum}

At finite frequencies, to have a theory consistent with current conservation it is essential to include the so called displacement currents \cite{Bla00}. This point is of vital importance to reproduce correctly the noise spectra measured experimentally. Although our discussion has focused, by construction,  on particle
currents so far, we show here how to include the effect of displacement
currents in our formalism. 

Let us illustrate this point by considering a quantum well with two terminals ($L$ and $R$) in contact with Fermi leads at different chemical potentials. There will be then a net current flowing through both terminals, but also, charge can ``accumulate'' in the well for some time. Therefore charge conservation can be expressed as
\beq
\dot{Q}(t) = I_L - I_R \equiv I_{accum},
\eeq
with $Q$ the charge in the well and $I_L$, $I_R$ referring to the currents through the left and right contacts respectively. $\dot{Q}$ represents the displacement current, $I_{dis}$, which can be partitioned as $I_{dis}=(\alpha+\beta)I_{dis}=I^R_{dis}+I^L_{dis}$, where $\alpha$ and $\beta$ describe how the displacement current is
divided between the right and left reservoirs (obviously
$\alpha+\beta=1$). This partitioning allows us to write the 
current conservation as $I_L-I^L_{dis}-(I_R+I^R_{dis})=0$.
Equivalently, $I_{tot}=I_L-I^L_{dis}=I_R+I^R_{dis}=\alpha I_L+\beta
I_R$, which is the so-called Ramo-Shockley theorem \footnote{In this paper we mean ``total'' cumulant when a subscript is ommited.}. In the simplest wide-band limit, the partition coefficients can be
written in terms of tunnel rates only as
$\alpha:=\Gamma_R/(\Gamma_L+\Gamma_R)$ and
$\beta:=\Gamma_L/(\Gamma_L+\Gamma_R)$ \cite{Guo99} \footnote{In a
Coulomb Blockade model, the partition coefficients read
$\alpha=C_R/(C_L+C_R)$ and $\beta=C_L/(C_L+C_R)$, where $C_L$, $C_R$
are the capacitances of each barrier and we have neglected
capacitive effects from the gate. See for instance
\cite{bruder-schoeller94}.}, and this will be the partitioning we will use throughout the paper. 

Experimentally, one can measure correlations of the current through
the device by transport measurements \cite{Reu03,Rez05}, or
indirectly by studying the current through a charge sensor, such as
a quantum point contact \cite{Lu03,Gus06,Fuj06}, that reveals
whether the well is ``charged'' or ``uncharged''. The second method
gives the statistics of the transport current only for very large
bias voltages (unidirectional counting) but, in general the
time-dependent transport current and the charge statistics are
different. Morevover, when the device itself is used as a detector, the
difference between transport fluctuations and charge fluctuations
leads to profound physical consequences. Unlike the inelastic
backaction induced by current fluctuations of the detector \cite{QPC3},
the one induced by charge fluctuations is the fundamental
Heisenberg backaction associated with the measurement \cite{Young09}.
Both transport and charge
fluctuations can be accounted for in our formalism by considering
counting fields
\beq
\chi_{tot} &:=& \chi_L + \chi_R, \label{chitot} \\
\chi_{accum} &:=& \beta \chi_L - \alpha \chi_R, \label{chiaccum} \eeq which lead to
respective jump operators \beq
\mathcal{J}_{\chi,tot}^{(n)}\equiv \alpha^n \mathcal{J}_{\chi,L}^{(n)} + \beta^n \mathcal{J}_{\chi,R}^{(n)}, \label{Jtot} \\
\mathcal{J}_{\chi,accum}^{(n)} \equiv \mathcal{J}_{\chi,L}^{(n)} +
(-1)^n \mathcal{J}_{\chi,R}^{(n)}, \label{Jaccum} \eeq where $\mathcal{J}_{\chi,L}$
and $\mathcal{J}_{\chi,R}$ refer to the two independent tunneling
processes occurring at each barrier. 

The ``total" cumulant through a two terminal device can be then calculated performing derivatives of the CGF with respect to $\chi_{tot}$ defined in (\ref{chitot}). This leads to expressions (\ref{currentformula})-(\ref{skewnessformula}) with ${\cal J}_{0}^{(n)}$ substituted by ${\cal J}_{0,tot}^{(n)}$ everywhere. Also, the spectrum of charge fluctuations
\beq
S^{(N)}_Q(\omega_1,\ldots,\omega_N):= \int_{-\infty}^\infty dt_1\ldots dt_N && e^{-i\omega_1 t_1}\ldots e^{-i\omega_Nt_N} \nonumber\\ &&\times {\cal T}_S \langle Q(t_1) \ldots
Q(t_N) \rangle_c,\eeq
follows from 
\beq
S^{(N)}_Q[\bm{\omega}] = \frac{\delta(i\omega_1+\ldots+i\omega_N)}{(i\omega_1)\ldots(i\omega_N)} S^{(N)}_{accum}[\bm{\omega}],
\eeq
with $S^{(N)}_{accum}[\bm{\omega}]$ obtained from eqs. (\ref{currentformula})-(\ref{skewnessformula}) upon the change ${\cal J}_0 \to {\cal J}_{0,acumm}$. For example, $S^{(2)}_Q(\omega) = (1/\omega^2) S^{(2)}_{accum}(\omega)$. Notice that for a capacitive conductor, due to the relation between charge and voltage, this charge noise is proportional to the voltage noise.
Finally the ``left'' and ``right'' cumulants can be computed with ${\cal J}_0 \to {\cal J}_{0,L}$ and ${\cal J}_0 \to {\cal J}_{0,R}$ respectively in (\ref{currentformula})-(\ref{skewnessformula}).

\section{Results} \label{Resultssec}

To illustrate our method, we analyze the transport
statistics of the prototypical example of spinless electrons passing
through a SRL model. The system consists on a two-terminal conductor with a discrete energy level in the central region, and is described
by the Hamiltonian \beq \label{SRLhamiltonian} \mathcal{H} =
\varepsilon {\ket 1}{\bra 1} + \sum_{k,\alpha\in L,R} \left(
\varepsilon_{k\alpha} + \mu_{\alpha} \right) c_{k\alpha}^{\dagger}
c_{k\alpha} + \sum_{k,\alpha\in L,R} {\cal V}_{k\alpha}
c_{k\alpha}^{\dagger} {\ket 0}{\bra 1} \; + \mathrm{H.c.}, \nonumber\\ \eeq where $k$ is
the momentum and ${\ket 0}$ and ${\ket 1}$ are the only two possible
states (referring to empty and occupied level) due to Coulomb blockade, with respective energies $0$ and $\varepsilon$. 

In the infinite bias limit (voltage $V$ much larger than the other
energy scales, excepting the bandwidth of the Fermi leads) the
Hamiltonian (\ref{SRLhamiltonian}) leads to the kernel
\beq\label{kernelinfty} \mathcal{W}(\chi) =
  \left(
  \begin{array}{cc}
    - \Gamma_L &\Gamma_R  e^{i\chi_R} \\
    \Gamma_L  e^{-i\chi_L} & -\Gamma_R
  \end{array}
  \right),
\eeq expressed in the basis $\lbrace {\ket 0}, {\ket 1} \rbrace$, and where $\Gamma_{\alpha} \approx \Gamma_{\alpha}(E) := \frac{2\pi}{\hbar} \sum_k |{\cal V}_{k\alpha}|^2 \delta(E-\varepsilon_{k\alpha})$ are the rates accounting for the system-reservoir coupling.
Using (\ref{currentformula}), (\ref{noiseformula}) and
(\ref{skewnessformula}), the simplicity of the model allows us to
derive analytic results in this limit; for example the current gives
$I_{stat}=\Gamma_L\Gamma_R/\Gamma$, where
$\Gamma:=\Gamma_L+\Gamma_R$, and the ``total'' noise expressed in
terms of the Fano factor ($F^{(2)}:=S^{(2)}/I_{stat}$) reads \beq
F^{(2)}(\omega) &=& \frac{ \Gamma_L^2 + \Gamma_R^2 +
(1-2\alpha\beta) \omega^2 }{\Gamma^2 + \omega^2}. \eeq

At finite bias voltages, the kernel in Eq. (\ref{kernelinfty}) is no longer valid. Among the various choices to calculate $\mathcal{W}(\chi)$ in this case, we use Schoeller's approach \cite{Schoeller09,Leijnse08,Emary09} (c.f. \ref{appKernel})
which allows us to calculate the Markovian kernel to the desired order (sequential tunneling in our case) without
further uncontrolled approximations (such as the secular approximation). It is important to mention that the frequency-dependent shot noise of the SRL model can be solved exactly \cite{Averin}, and therefore one does not need to use the above approximations. However, to the best of our knowledge, a finite-frequency study for this model beyond shot noise is yet lacking. Here we use the exact solution as a benchmark of the Markovian approximation in order to identify the regions of validity of our theory. This benchmark is important because most of the papers in the literature discussing shot noise in the context of QMEs make use of the Markovian approximation.

Another important check for the theory is to reproduce the fluctuation-dissipation theorem (FDT) in the appropriate regimes. Near linear response, that is, for applied voltages $V$ much smaller than the temperature $T$, the low-frequency noise spectrum should follow the Jonhson-Nyquist relation $S^{(2)} = 2kTG$ \cite{Johnson28, Nyquist28}, where $G$ is the dc conductance. This equilibrium FDT was later extended by Callen and Welton to include quantum fluctuations \cite{Call51}, relevant when the measured frequencies are larger than the temperature. The FDT takes then the form $S^{(2)}(\omega) = \hbar\omega \mathrm{coth}(\frac{\hbar\omega}{2kT}) G(\omega)$, where $G(\omega)$ is the ac conductance. This expression can be equivalently written in terms of the Bose-Einstein distribution $N_{\mathrm{BE}} (\omega) \equiv 1/[e^{\frac{\hbar\omega}{kT}}-1]$, since $\mathrm{coth}(\frac{\hbar\omega}{2kT})=2N_{\mathrm{BE}}(\omega)+1=N_{\mathrm{BE}}(\omega)-N_{\mathrm{BE}}(-\omega)$, and it becomes clear that the \textit{symmetrized} noise, which we are considering here, contains both absorption and emission.
Out of equilibrium, a fluctuation-dissipation relation can be also found for some particular cases, such as tunnel junctions \cite{Dahm69,rogovin-scalapino} or for quantum dots in the weak cotunneling regime \cite{Suk01}. For a two-terminal conductor driven out of equilibrium, the \textit{symmetrized} \footnote{For a \textit{non-symmetrized} version of the noise spectrum through a two-terminal conductor c.f. \cite{QPC3}.} noise spectrum takes the general form \cite{QPC1, QPC2, freq-exp3, Bla00}
\beq \label{FDTfull}
S^{(2)}(\omega) &=& \hbar\omega \mathrm{coth}\left(\frac{\hbar\omega}{2kT}\right) \sum_n D_n^2 + \left[ \frac{(\hbar\omega+eV)}{2} \mathrm{coth}\left(\frac{\hbar\omega+eV}{2kT}\right) \right. \nonumber \\ && \left. + \frac{(\hbar\omega-eV)}{2} \mathrm{coth}\left(\frac{\hbar\omega-eV}{2kT}\right) \right] \sum_n D_n(1-D_n),
\eeq
where $D_n$ is the transmission coefficient of the conduction channel $n$. This expression has various interesting limits. First, in the tunneling regime ($D_n\ll1$), it gives the non-equilibrium fluctuation-dissipation theorem (NEFDT) as reported in \cite{Dahm69, rogovin-scalapino} for tunnel junctions:
\beq \label{FDTfreq} 
S^{(2)}(\omega)=\frac{1}{2}\sum_{p=\pm} I_{stat}(eV+p\hbar\omega)\mathrm{coth}\left(\frac{eV+p\hbar\omega}{2kT}\right),
\eeq
expression that is also exact for quantum dots in the weak cotunneling regime \cite{Suk01}, and whose zero-frequency limit $S^{(2)}= I_{stat} \mathrm{coth}(\frac{eV}{2kT})$ has been derived in the context of counting statistics \cite{Lev04}. For low voltages, $eV\ll kT$, equation (\ref{FDTfull}) recovers the Callen and Welton equilibrium relation, and if also $\hbar\omega\ll kT$, it gives the Johnson-Nyquist FDT (thermal noise regime). Finally, if $eV\gg kT,\;\hbar\omega$ (shot noise regime), we find $S^{(2)}=\zeta I$, where the coefficient $\zeta\equiv \sum_n D_n (1-D_n) / \sum_n D_n$ is the Fano-factor.

\begin{figure}
 \begin{center}
\epsfig{file=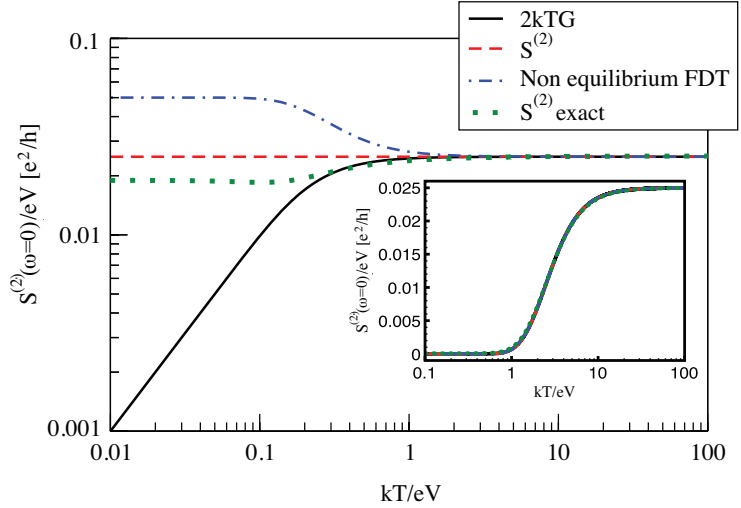, clip=true, angle=0, width=0.75\linewidth} 
 \caption{Zero-frequency noise $S^{(2)}(\omega=0)$ at finite voltage for the single resonant level model as a function of temperature. For comparison we also show the FDT, the NEFDT, and the exact solution. 
  In the main figure the level is within the bias window $\varepsilon=0$. The inset  shows a regime with the level outside the bias window $\varepsilon/eV=5$.
  Rest of parameters: $\Gamma_L/eV=\Gamma_R/eV=0.1$.  In the main figure all the quantities coincide when $kT\geq eV$, as expected. 
  In the inset all fluctuations are thermal and therefore, all quantities coincide in the whole range of temperatures. The typical physical units are $T\sim 10-100\;\mathrm{mK}$, $V\sim 10-100\;\mathrm{\mu V}$, $\Gamma\sim 10-100\;\mathrm{MHz}$.
   \label{fig1}
}
 \end{center}
\end{figure}

In this section we show the solution given by our theory for the SRL model, as well as how it recovers the the FDT and the NEFDT in the appropriate limits. By contrasting these results with the exact solution, we will be able to show that the Markovian approximation does not contain quantum fluctuations, thereby needing a non-Markovian approach to capture the physics of quantum noise \cite{Marcos10}.
It is therefore interesting to see to what extend the four results coincide, and in what regimes our Markovian approach is valid and recovers the proper physics. We will see that when $kT\gg eV,\; \hbar\omega$, the theory captures well both the exact solution and the FDT and NEFDT. Also, in transport configurations, with the level within the bias window, the Markovian approximation agrees well with the exact solution, reproducing in particular previous studies with $eV\gg kT$ \cite{Gurvitz96, Aguado04, Braun-etal, Choi}. However, in a situation with level outside the bias window, the Markovian approach presented here does not capture quantum noise physics, effect that we observe at high frequencies ($\hbar\omega\gtrsim kT, \; eV$). Although in this situation transport due to cotunneling processes becomes more relevant, the difference is due to the Markovian assumption as can be seen using a non-Markovian extension of the theory \cite{Marcos10}.
We also study the finite-frequency skewness of the current distribution as given by equation (\ref{skewnessformula}). This shows to be insensitive to thermal fluctuations near equilibrium, therefore revealing the ``shot'' contribution in a situation in which thermal fluctuations dominate in the noise spectrum ($kT \gg eV, \; \hbar\omega$). 

\begin{figure}
 \begin{center}
  \epsfig{file=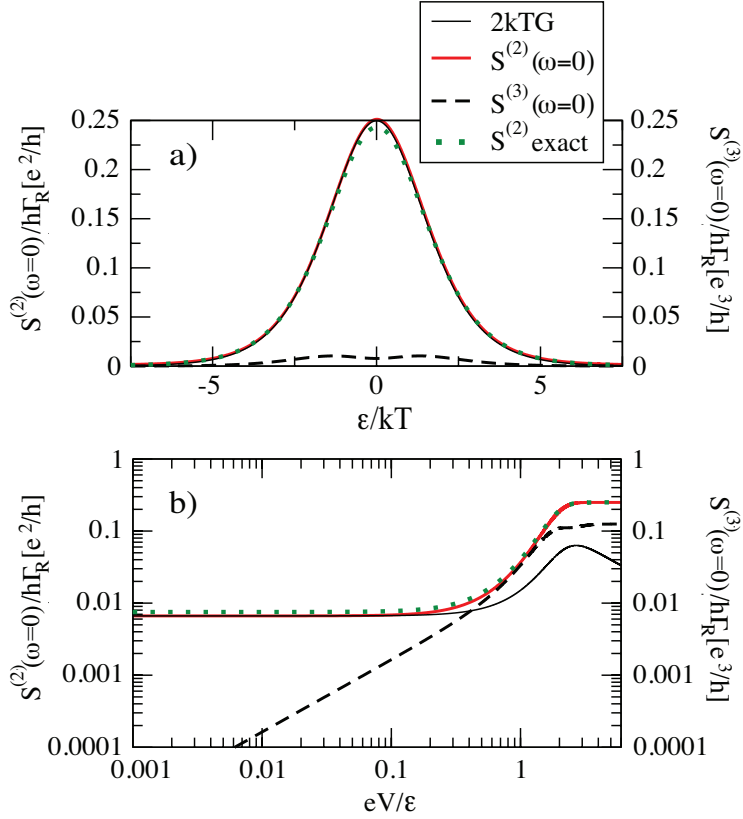, clip=true, angle=0, width=0.75\linewidth} 
 \caption{a) Noise and skewness near linear response, for $h\Gamma_L/kT=h\Gamma_R/kT=eV/kT=0.05$, as a function of $\varepsilon$. b) Noise and skewness as a function of voltage. We also show the equilibrium fluctuation-dissipation expression $2kTG$ and the exact solution for comparison.
   \label{fig2}
}
 \end{center}
\end{figure}

\subsection{Zero-frequency counting statistics.} \label{FCSresults}

Let us start by showing the zero-frequency noise spectra corresponding to the SRL model. Although this limit has already been studied in detail, a full comparison between our theory and the exact solution will help to understand the finite-frequency case. Particularly important is the linear response regime, at which studies of this model are scarce. As mentioned before, in this regime the noise should exhibit thermal fluctuations in order to fulfill a fluctuation-dissipation
relation, while the skewness, on the other hand, should go to zero as the voltage $V$ goes to zero \cite{Lev04}.
Fig.~\ref{fig1} shows how our calculation captures correctly the
FDT, $S^{(2)}=2kTG$, in the proximity of linear response, $kT
\gtrsim eV$. For comparison, we also plot the zero-frequency limit of the NEFDT in Eq. (\ref{FDTfreq}), namely
$S^{(2)}= I_{stat} \mathrm{coth}(\frac{eV}{2kT})$ \cite{Lev04}. 
In the opposite regime, $kT\lesssim eV$, the Markovian approximation is larger than the exact solution, discrepancy that can be understood as originated from the lack of cotunneling contributions in our calculation \cite{Emary09}. As expected, below $kT/eV\sim1$ the FDT is not fulfilled. We can also see that the NEFDT, exact for tunnel junctions, for a two-terminal device performs quite badly when $kT\lesssim eV$, but correctly in the opposite limit. This failure of the NEFDT at low temperatures disappears when the level is outside the bias window. This is a low-current regime, and thus a tunneling limit. The inset of Fig.~\ref{fig1} shows this situation, where all fluctuations are thermal and the four curves coincide in the whole range of temperatures. At finite frequencies we will expect a quantum noise step in the spectrum at frequencies $\hbar\omega\sim\varepsilon$, effect that will be studied in the next subsection.

\begin{figure}
 \begin{center}
  \epsfig{file=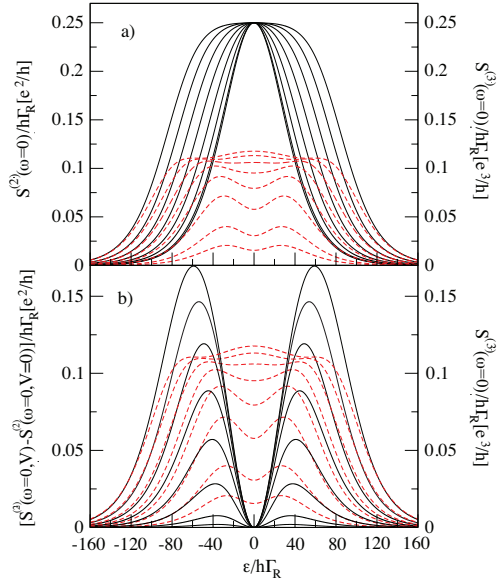, clip=true, angle=0, width=0.5\linewidth} 
 \caption{a) Increasing noise (black full lines going up) and skewness (red dashed lines going up) as a function of $\varepsilon$ for increasing voltages $eV/h\Gamma_R=1,10,20,40,60,80,100,120$, (parameters $kT/h\Gamma_R=20, \Gamma_L/\Gamma_R=1$). b) Excess noise  $S^{(2)}(V)-S^{(2)}(V=0)$ and skewness versus $\varepsilon$ for increasing voltages.
   \label{fig3}
}
 \end{center}
\end{figure}

The behaviour of the zero-frequency noise spectrum close to equilibrium with respect to $\varepsilon$ is shown in Fig.~\ref{fig2}a. Here we see how the FDT is fulfilled by our theory and the good agreement with the exact solution. We also plot the zero-frequency skewness, that although of small magnitude in the same scale, is nonzero in a situation where the noise spectrum is completely dominated by thermal fluctuations. This insensitivity of the skewness to temperature allows us to extract intrinsic correlation effects in near-equilibrium conditions.
In Fig.~\ref{fig2}b we show the same quantities as a function of voltage. Interestingly, the Markovian result coincides with the exact solution in the whole range of voltages. The FDT, however, starts to disagree with these for voltages $eV/\varepsilon\gtrsim 0.2$. As anticipated, the skewness vanishes as the voltage goes to zero.
In Fig.~\ref{fig3}a  we plot noise and skewness as a function of $\varepsilon$ for increasing voltages. As the bias increases, the skewness (dashed lines) shows peaks evolving into plateaus at values of $\varepsilon$ corresponding to the chemical potentials of the reservoirs. This effect, which is due to non-equilibrium fluctuations, is completely masked in the noise (solid lines) even at the highest voltages due to thermal fluctuations. This is clearly seen in Fig.~\ref{fig3}b, where we show the same comparison after substracting thermal fluctuations to the noise value (excess noise defined as $S^{(2)}(V)-S^{(2)}(V=0)$). Here it is clear that at low detuning, $\varepsilon\lesssim eV/2$ (position of the peaks in the figure), and when $kT\gtrsim eV$, the skewness can reveal the ``shot'' contribution, while this is masked by thermal fluctuations in the noise spectrum.

\subsection{Finite-frequency counting statistics.} \label{FCSwresults}

 \begin{figure}
 \begin{center}
  \epsfig{file=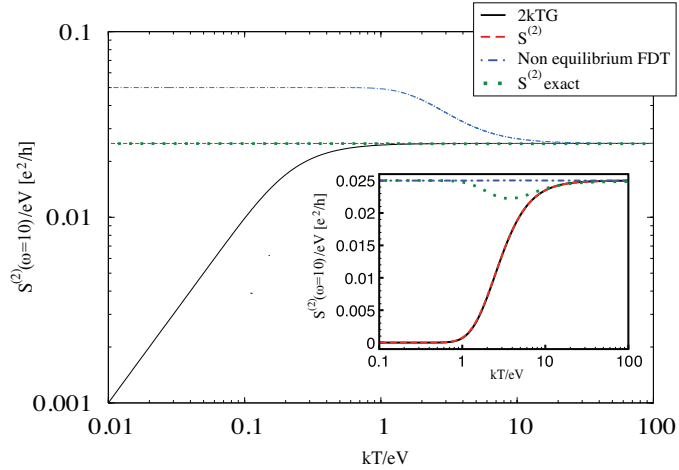, clip=true, angle=0, width=0.75\linewidth} 
 \caption{Noise, FDT and NEFDT as a function of temperature for $\omega/eV=10$.
  In the main figure the level is within the bias window $\varepsilon=0$. The inset  shows a regime with the level outside the bias window $\varepsilon/eV=5$.
  Rest of parameters: $h\Gamma_L/eV=h\Gamma_R/eV=0.1$.  In the main figure all quantities coincide when $kT\geq eV+\hbar\omega$.
  In the inset all fluctuations are thermal and, therefore, the shot noise and the FDT coincide in the whole range of temperatures. When $kT\leq eV+\hbar\omega$, the NEFDT is above due to quantum fluctuations. The exact solution contains also corrections due to cotunneling processes, which are dominating in this regime.
   \label{fig4}
}
 \end{center}
\end{figure}

To study the case of finite frequencies, we use our formulae (\ref{currentformula})-(\ref{skewnessformula}) applied to the SRL model. In a situation with the level within the bias window, we find a similar behaviour to Fig.~\ref{fig1}. However, now the NEFDT -- equation (\ref{FDTfreq}) -- is fulfilled for temperatures $kT\gtrsim eV+\hbar\omega$. This is shown in Fig.~\ref{fig4}. Remarkably, at finite frequencies the Markovian approximation is basically exact in this direct transport regime.
In the high-bias regime $eV\gg \hbar\omega, \; kT$ we also find that the Markovian approximation agrees perfectly with the exact result (not shown), in accordance with previous studies \cite{Gurvitz96, Aguado04, Braun-etal, Choi}. When the level lies outside the voltage window, the situation changes drastically (see inset of Fig.~\ref{fig4}). Here the Markovian approximation is no longer appropriate when $kT\lesssim eV+\hbar\omega$. Both the exact result and the NEFDT contain quantum fluctuations, while the Markovian calculation only captures the thermal contribution (and therefore fulfills the FDT). The exact solution presents a small structure at temperatures of the order of $\varepsilon$. This cannot be resolved with the NEFDT. As expected, when $kT\gtrsim eV+\hbar\omega$, all curves coincide.

 \begin{figure}
 \begin{center}
  \epsfig{file=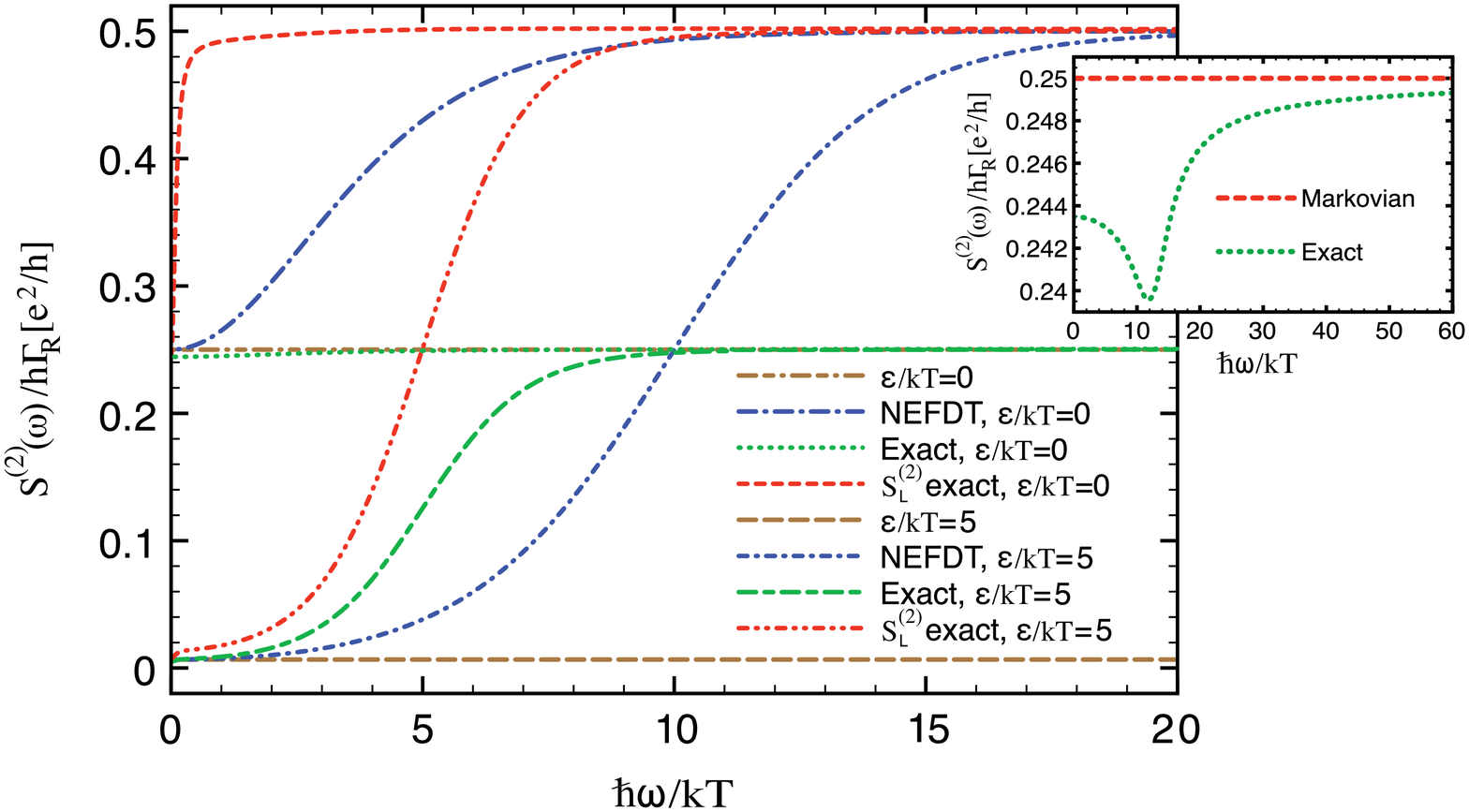, clip=true, angle=0, width=0.75\linewidth} 
 \caption{$S^{(2)}$ near linear response ($eV/kT=0.0005$) as a function of frequency. For comparison we also show the NEFDT (Eq.~(\ref{FDTfreq})) and the exact solution. $S^{(2)}(\omega)$ is flat for the whole range of frequencies, and coincides with the equilibrium FDT as expected. The NEFDT however disagrees with these two, showing also quantum fluctuations which are absent in the Markovian noise spectrum. The quantum noise steps shown by the NEFDT are however at $\hbar\omega=2\varepsilon$, in contrast to the exact solution, which shows steps at $\hbar\omega=\varepsilon$. This is due to the fact that the NEFDT works well for tunnel junctions, but does not capture partition noise. This becomes clear also from the saturation value at large frequencies, as described in the text. Rest of parameters: $h\Gamma_L/kT=h\Gamma_R/kT=0.05$. The inset compares the exact solution with the Markovian approximation for a different regime, namely $eV/kT = 25$. We see that while the Markovian limit is flat for all frequencies, the exact solution presents a dip at $\hbar\omega=\pm|\varepsilon \pm eV/2|$. Rest of parameters: $\varepsilon=0, h\Gamma_L/kT=h\Gamma_R/kT=0.25$.
\label{fig5}
}
 \end{center}
\end{figure}

The general trends explained so far become more evident in Fig.~\ref{fig5}, where we plot noise and the fluctuation-dissipation theorem near linear response as a function of frequency. Here the Markovian noise is always flat and equals $S^{(2)}=2kTG$, whereas the NEFDT and the exact solution lie above and show quantum noise steps. 
Let us start by considering the case $\varepsilon=0$. In the whole range of frequencies, the Markovian approximation is basically exact in this situation of direct transport. The NEFDT shows the correct zero-frequency limit, since then the fluctuations are purely thermal. At high frequencies, however, the NEFDT converges to the Poisson value of a single barrier with tunneling rate $\bar{\Gamma}/2$, being $\bar{\Gamma}:=(\Gamma_L+\Gamma_R)/2$, (c.f. $S_L^{(2)}$ in the plot). This is in agreement with the validity of equation (\ref{FDTfreq}) for a tunnel junction, and in contrast with the exact solution, which contains partitioning, and therefore its $\omega\to\infty$ limit is $\bar{\Gamma}/4$. In the case in which the energy lies outside the bias window ($\varepsilon/kT=5$ in the figure), transport is possible because of the finite temperature as well as due to quantum fluctuations. The Markovian noise only contains the former and is flat with frequency, while the exact result contains both and shows a quantum noise step centered at $\hbar\omega/kT=\varepsilon/kT=5$. Although in this regime cotunneling contributions are important, the difference lies in the Markovian approximation, as can be seen using a non-Markovian theory \cite{Marcos10}. The NEFDT shows in this situation a quantum noise step centered at $\hbar\omega=2\varepsilon$. This discrepancy with respect to the exact solution can be understood in terms of the tunneling approximation leading to Eq. (\ref{FDTfreq}), which presents a step located at an effective chemical potential $2\varepsilon$. Again, the high frequency limit coincides with that of $S_L^{(2)}$.
The intermediate regime where $\hbar\omega \sim eV$ is studied in the inset. Here, we set $eV/kT=25$ and observe a flat behaviour for the Markovian solution, whereas the exact solution presents a dip at $\hbar\omega = \pm | \varepsilon \pm eV/2|$ (coinciding with the position of the chemical potentials with respect to the energy level). This clearly illustrates how the Markovian approximation captures well the physics in the linear response regime and a direct transport configuration, but when the frequency is comparable to the applied bias, it fails to capture the quantum noise.

 \begin{figure}
 \begin{center}
  \epsfig{file=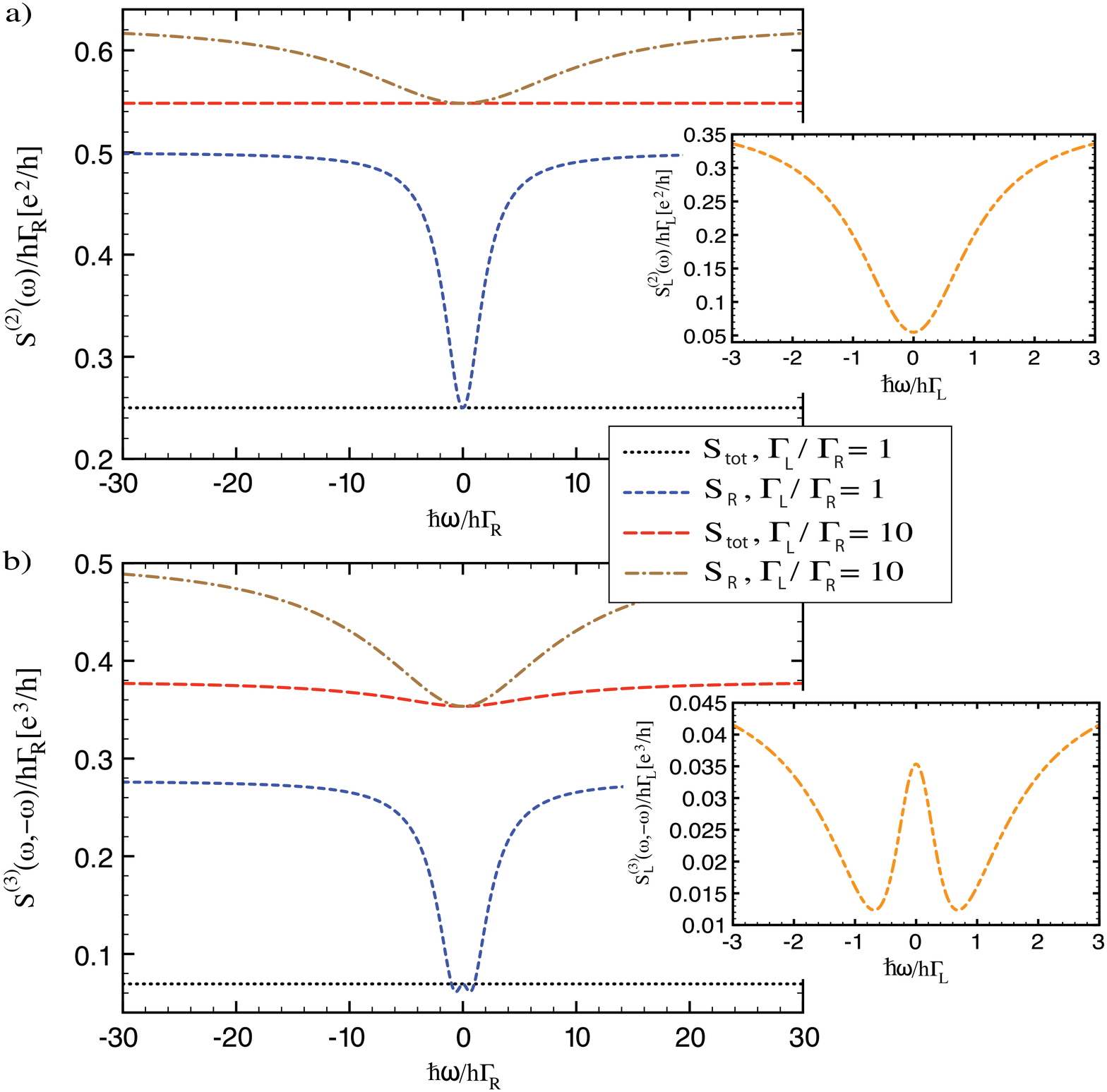, clip=true, angle=0, width=0.75\linewidth} 
 \caption{a) $S^{(2)}(\omega)$ and b) $S^{(3)}(\omega,-\omega)$ for $eV/h\Gamma_R=50$, $kT/h\Gamma_R=20$, and $\varepsilon=0$. The spectra for particle currents and for total currents significantly deviate form each other, even for large asymmetry. The insets correspond to noise and skewness through the left barrier for $\Gamma_L/\Gamma_R=10$.
   \label{fig6}
}
 \end{center}
\end{figure}

We now proceed to discuss the finite-frequency noise and skewness spectra of the total- and particle- current distribution. 
In the previous discussion, the Markovian noise was always flat as a function of frequency, something which is well known for symmetric systems ($\Gamma_L=\Gamma_R$) -- see for instance \cite{Bla00}. In the case $\Gamma_L\neq\Gamma_R$, a proper partitioning of displacement currents (see discussion in subsection \ref{total}) becomes essential as we will show next, and the way this is made affects significantly the spectrum. 
Fig.~\ref{fig6}a shows $S^{(2)}(\omega)$ in a transport configuration, $eV/h\Gamma_R=50$ and $\varepsilon=0$. As in the results shown previously, the total noise spectrum is flat for a symmetric configuration. Interestingly, this flat behaviour persists even when the system is made asymmetric ($\Gamma_L\neq\Gamma_R$). This is due to the current-partitioning model assumed here: $I_{tot} = \alpha I_L + \beta I_R$ with $\alpha = \Gamma_R/\Gamma$ and $\beta = \Gamma_L/\Gamma$. The noise spectrum corresponding to particle currents displays information about the rates; in contrast to the total noise, it shows a dip with half-width $2\Gamma$. In Fig.~\ref{fig6}b we show the skewness along the representative direction $\omega'=-\omega$. Interestingly, the skewness corresponding to the total current starts to develop a dip that shows the asymmetry of the system. The particle-current skewness presents a similar behaviour to the noise counterpart. However, for $(3-\sqrt{5})/2\leqslant \Gamma_L/\Gamma_R\leqslant (3+\sqrt{5})/2$ it develops a minimum whose position depends on the value of the rates. In the asymmetric case, $\Gamma_L\neq\Gamma_R$, the particle-current noise and skewness can even present different lineshapes. This can be seen contrasting the insets of Fig.~\ref{fig6}a and Fig.~\ref{fig6}b. In the linear response regime the curves for the noise look similar (not shown). The skewness on the contrary goes to zero in magnitude and shows a structure that depends on temperature, and that changes from a dip to a peak as $\varepsilon$ is increased from zero to a finite value. In summary, we see that the spectra for total and particle currents differ significantly from each other even for large asymmetry. This means that the assumption of calculating noise spectra using particle currents only, used commonly in the literature, is flawed. Here we have assumed the current partitioning given by $\alpha=\Gamma_R/\Gamma$, $\beta=\Gamma_L/\Gamma$. If the more simplistic partitioning $\alpha=\beta=1/2$ is assumed, the results for the total cumulants in the asymmetric case change drastically (not shown). In particular, the noise is then no longer flat but has a dip structure, and the skewness shows a peak around zero frequency.

\section{Conclusions} \label{ConclusionsSec}
In conclusion, we have developed a theory of frequency-dependent counting statistics of electron transport through nanostructures within the framework of Markovian quantum master equations. We have illustrated our method with calculations of noise and skewness in a single resonant level model at finite bias voltages and frequencies. By comparing with both the exact solution and the finite-frequency version of the nonequilibrium fluctuation-dissipation theorem, Eq.~(\ref{FDTfreq}), we have identified the regimes of validity of our Markovian theory at finite frequencies. In particular, we have shown that the Markovian limit is basically exact in transport configurations (level within the bias voltage window), as long as $\hbar\omega \gg eV$ or $\hbar\omega\ll eV$. In intermediate situations where $\hbar\omega \sim eV$, or with the level outside the bias window, the Markovian limit fails  at finite frequencies due to the lack of quantum fluctuations \cite{Marcos10}. 

We have also discussed how the noise spectra for particle currents and for total currents significantly deviate from each other, even for large asymmetries $\Gamma_R/\Gamma_L \neq 1$. This demonstrates that calculating spectra using particle currents only leads to incorrect results in general.
Our method allows the calculation of finite-frequency current cumulants of arbitrary order, as we have explicitly shown for the second and third order cumulants, Eqs. (\ref{noiseformula}) and (\ref{skewnessformula}). These formulae generalize previous zero-frequency expressions and can be viewed as an extension of MacDonald's formula beyond shot noise. Recently, this has been extended to study the time-averaged shot noise spectrum in the presence of periodic ac fields \cite{Clerk-Girvin,Wu-Timm}. Interesting extensions of our study along these lines would allow us to study frequency-dependent high-order cumulants  in nanostructures driven by time-dependent fields, or, even more challenging, in systems showing nontrivial non-linear dynamics such as self-sustained oscillations without external time-dependent driving \cite{Self-sustained}.

\section*{Acknowledgments}

We thank C. Flindt for stimulating discussions, for his help in recovering the proper zero-frequency limit of our theory, and for careful reading of the manuscript. We also thank A. Braggio, M. B\"uttiker, T. Novotn\'y, M. Wegewijs, S. Kohler, J. L. F. Barb\'on, M. Garc\'ia P\'erez, B. W\"unsch, and P. Zedler for discussions. Work supported by MEC-Spain (Grant No. FIS2009-08744)  and Acci\'on Integrada Spain-Germany Grant No. HA2007-0086.  D. M. acknowledges funding from grant  FPU AP2005-0720.

\appendix

\section{Derivation of frequency-dependent cumulants}
\label{appDiagrams}

The expressions (\ref{currentformula})-(\ref{skewnessformula}) follow from derivatives of moment generating functions. Performing derivatives of (\ref{GN}) we find
\beq \label{Igreater}
\langle I(z) \rangle^{>} =z\partial_{\chi} \Big\langle \Omega(\chi, z) \Big\rangle\Big\vert_0= z^{-1}\langle {\cal J}_0^{(1)} \rangle \eeq
\beq \label{S2greater}
S_m^{(2)>}(z_1,z_2) &=& \left(z_1z_2\right) \left.
   \partial_{\chi_1} \partial_{\chi_2} \Big\langle
   \Omega(\chi_2,z_2)\Omega(\chi_{12},z_{12})
   \Big\rangle
   \right|_0 \nonumber \\
    &=&\left(z_1z_2\right) z_2^{-1}z_{12}^{-2} \langle 2 {\cal J}_0^{(1)} \Omega_0(z_{12}) {\cal J}_0^{(2)} \nonumber \\ &&+ z_{12} {\cal J}_0^{(2)} \Omega_0(z_2) \Omega_0(z_{12}) {\cal J}_0^{(1)} + {\cal J}_0^{(2)} \rangle \eeq
\beq \label{S3greater}
S_m^{(3)>}(z_1,z_2,z_3) &=&  \left(z_1z_2z_3\right) \left.
   \partial_{\chi_1}\partial_{\chi_2}\partial_{\chi_3}
    \Big\langle
    \Omega(\chi_3,z_3)
    \Omega(\chi_{23},z_{23})
    \Omega(\chi_{123},z_{123})
	  \Big\rangle
   \right|_0 \nonumber \\
	  &=& \left(z_1z_2z_3\right) z_3^{-1} z_{123}^{-1} \nonumber \\  
	&&\times \langle
		  {\cal J}_0^{(1)} \Omega_0(z_{3}) \Omega_0(z_{23})
		  {\cal J}_0^{(1)} \Omega_0(z_{23})\Omega_0(z_{123}){\cal J}_0^{(1)} \nonumber \\
		  &&+ 2{\cal J}_0^{(1)} \Omega_0(z_{3}) \Omega_0(z_{23})\Omega_0(z_{123})
		  {\cal J}_0^{(1)} \Omega_0(z_{123}){\cal J}_0^{(1)}
	  \nonumber\\ 
		  &&+4 z_{23}^{-1} {\cal J}_0^{(1)} \Omega_0(z_{23}) \Omega_0(z_{123})
		  {\cal J}_0^{(1)}\Omega_0(z_{123}){\cal J}_0^{(1)}
	  \nonumber\\ 
		&&
		  + 2 z_{23}^{-1} {\cal J}_0^{(1)} \Omega_0(z_{23})
		  {\cal J}_0^{(1)} \Omega_0(z_{23})\Omega_0(z_{123}){\cal J}_0^{(1)} \nonumber \\	
		&&
		  +6 z_{23}^{-1} z_{123}^{-1} {\cal J}_0^{(1)}  \Omega_0(z_{123})
		  {\cal J}_0^{(1)} \Omega_0(z_{123}){\cal J}_0^{(1)} \nonumber \\		    	  
	    &&+ {\cal J}_0^{(1)}  \Omega_0(z_{3})\Omega_0(z_{23})\Omega_0(z_{123}){\cal J}_0^{(2)} \nonumber \\
	    &&+2 z_{23}^{-1} {\cal J}_0^{(1)}  \Omega_0(z_{23})\Omega_0(z_{123}){\cal J}_0^{(2)}
	    	  \nonumber\\ 
		  &&+ z_{23}^{-1} {\cal J}_0^{(2)}  \Omega_0(z_{23})\Omega_0(z_{123}){\cal J}_0^{(1)} \nonumber\\	
		  &&+3 z_{23}^{-1} z_{123}^{-1} {\cal J}_0^{(2)}  \Omega_0(z_{123}){\cal J}_0^{(1)}
	  \nonumber\\ 		  	  
		&&
	    +3 z_{23}^{-1} z_{123}^{-1} {\cal J}_0^{(1)}  \Omega_0(z_{123}){\cal J}_0^{(2)} \nonumber\\
		  &&+z_{23}^{-1} z_{123}^{-1}{\cal J}_0^{(3)}
	  \rangle
	  ,
\eeq
where $z_{ij}:=z_i+z_j$, $z_{ijk}:=z_i+z_j+z_k$. Next we use
\beq \label{simplificationOmegas}
\Omega_0(z_2)\Omega_0(z_1+z_2) = \frac{1}{z_1} \Big[ \Omega_0(z_2) - \Omega_0(z_1+z_2) \Big],
\eeq
and add the ``lesser" part ($<$) corresponding to negative Laplace frequencies. At this point we change to ``physical'' frequencies $\omega:=\omega_2+\ldots+\omega_N$, $\omega':=\omega_3+\ldots+\omega_N$, etc., and symmetrize the result. This means adding the expressions corresponding to all the possible frequency switchings. In this step we take into account that 
\beq \label{deltaIdentity}
\lim_{\eta\to 0} \left(\frac{1}{i\omega + \eta} + \frac{1}{-i\omega+\eta} \right)= \lim_{\eta\to 0} \frac{2\eta}{\omega^2+\eta^2} = 2\pi\delta(\omega),
\eeq 
where $\eta\rightarrow 0$ is a small parameter coming from the ``greater" ($>$) or ``lesser" ($<$) parts. Finally, we make use of this energy conservation inherited from the time-translational symmetry of the cumulants. We then arrive to the equations (\ref{currentformula})-(\ref{skewnessformula}) used in the main text.
Importantly, after frequency symmetrization one can realize that the first three cumulant formulae are equal to their moment counterparts.

\subsection*{A.1. Diagrams.}

\begin{figure}[p]
 \begin{center}
 \epsfig{file=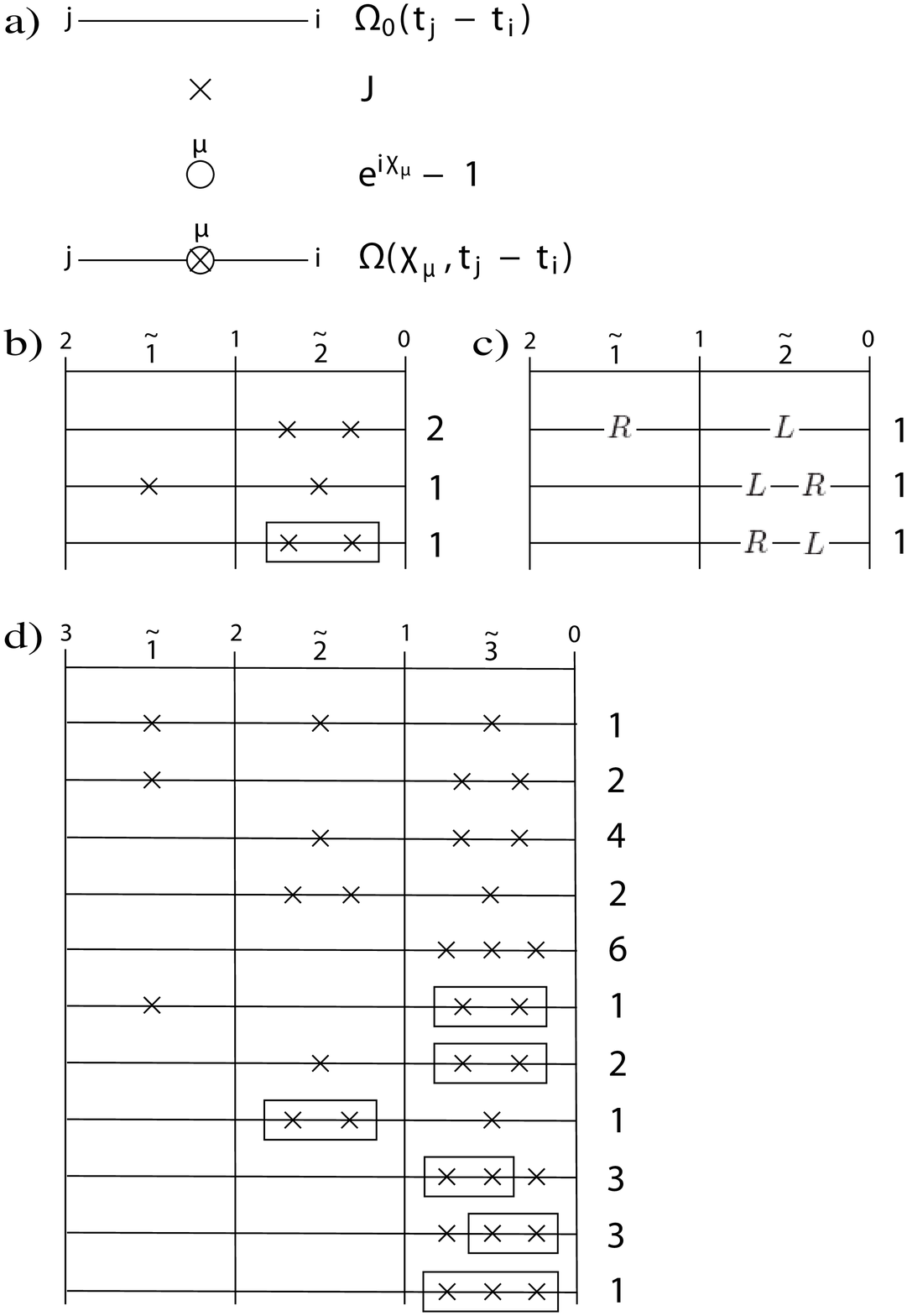, clip=true,width=1\linewidth}
  \caption{Diagrammatics to obtain the frequency-dependent cumulants.
     {\bf a)} Building pieces for the diagrammatic technique. A line is associated to a bare propagator, a cross with a jump operator, and a circle with the time dependence of ${\cal J}$ (term $(e^{i\chi} - 1)$ in the single-particle unidirectional tunneling case). Derivatives with respect to counting fields eliminate circles correspondingly. Diagrams can be simplified using rules like (\ref{simpDiagrams}).
    {\bf b)} Diagrams for the noise. Reading this table we find the expression $2\Omega_0(\tilde{z}_1)\Omega_0(\tilde{z}_2){\cal J}^{(1)}_0\Omega_0(\tilde{z}_2){\cal J}^{(1)}_0\Omega_0(\tilde{z}_2) + \Omega_0(\tilde{z}_1){\cal J}^{(1)}_0\Omega_0(\tilde{z}_1)\Omega_0(\tilde{z}_2){\cal J}^{(1)}_0\Omega_0(\tilde{z}_2) + \Omega_0(\tilde{z}_1)\Omega_0(\tilde{z}_2){\cal J}^{(2)}_0\Omega_0(\tilde{z}_2)$.
    {\bf c)} Diagrams for the second order cross correlation $S^{(2)}_{LR}$. Reading this table we find the expression $\Omega_0(\tilde{z}_1){\cal J}^{(1)}_{0,R}\Omega_0(\tilde{z}_1)\Omega_0(\tilde{z}_2){\cal J}^{(1)}_{0,L}\Omega_0(\tilde{z}_2) + \Omega_0(\tilde{z}_1)\Omega_0(\tilde{z}_2){\cal J}^{(1)}_{0,L}\Omega_0(\tilde{z}_2){\cal J}^{(1)}_{0,R}\Omega_0(\tilde{z}_2) + \Omega_0(\tilde{z}_1)\Omega_0(\tilde{z}_2){\cal J}^{(1)}_{0,R}\Omega_0(\tilde{z}_2){\cal J}^{(1)}_{0,L}\Omega_0(\tilde{z}_2)$.
    {\bf d)} Diagrams to derive the frequency-dependent skewness formula.
  \label{diagrams}
}
 \end{center}
\end{figure}

Interestingly, the results (\ref{Igreater})-(\ref{S3greater}) given above can be derived following a diagrammatic technique, similarly to how this is done with Feynman diagrams in the expansion of the partition function or the $S$-matrix. Similarly we write the CGF in terms of a series expansion, either in the time domain or in the frequency space. To that end we expand each of the $\chi$-dependent propagators in the CGF as a Dyson series:
\beq\Omega(\chi,z) =
 \frac{1}{z-\mathcal{W}(\chi)}
=\Omega_0(z)\sum_{n=0}^{\infty} \left[
\mathcal{J}_{\chi} \Omega_0(z) \right]^n.
\eeq
This suggests the use of diagrams of the form given in Fig.~\ref{diagrams} (a).
In the frequency domain \footnote{If we work in the time domain, it is enough to label the propagating lines with the corresponding times at the beginning and end of each line (see Fig.~\ref{diagrams} (a)).} this rules are:
\begin{itemize}
\item To each bare propagator $\Omega_0(\tilde{z}_k)$ in the expansion we associate a line with a superscript $\tilde{k}\equiv \sum_{i=N+1-k}^N i$, where $N$ is the order of the cumulant we want to obtain.
\item To each jump operator ${\cal J}_{\tilde{\chi}_k}$ in the expansion we associate an encircled cross with superscript $\tilde{k}$.
\end{itemize}
The formula for the generating function to a given order can therefore be written diagrammatically. For example, to second order we have 
\beq
{\cal G}(\bm{\chi},\bm{z}) &=& {\cal T}_S Tr\{\Omega(\tilde{\chi}_1,\tilde{z}_1)\Omega(\tilde{\chi}_2,\tilde{z}_2)\rho_\mathrm{S}^{stat}\} \nonumber\\ &=& {\cal T}_S Tr\{ \left( \Omega_0(z_2) + \Omega_0(z_2){\cal J}_{\chi_2} \Omega_0(z_2) + \ldots \right) \nonumber\\ &\times& \left( \Omega_0(z_1+z_2) + \Omega_0(z_1+z_2){\cal J}_{\chi_1+\chi_2} \Omega_0(z_1+z_2) + \ldots \right) \rho_\mathrm{S}^{stat} \}. \nonumber\\
\eeq
We can then multiply the different terms using diagrams as described above. The multiplication of propagators implies joining them together. The result can be simplified using the property ${\cal J}_{\chi_1+\chi_2}={\cal J}_{\chi_1}{\cal J}_{\chi_2}+{\cal J}_{\chi_1}+{\cal J}_{\chi_2}$, which diagrammatically is denoted as
\beq \label{simpDiagrams}
\bigotimes^{12}=\bigotimes^1\bigotimes^2+\bigotimes^1+\bigotimes^2.
\eeq
Here, the super-index $12$ denotes an associated frequency $z_1+z_2$ and counting field $\chi_1+\chi_2$. 
Next, to arrive to the frequency-dependent \textit{moment}, we take derivatives with respect to counting fields. Diagrammatically, the derivative $\partial_{\chi_k}$ means removing a circle with index $k$. From here we can rewrite the expression analytically. The outcome can be simplified using (\ref{simplificationOmegas}), and needs to be multiplied by $z_1z_2$ (case $N=2$), coming from the Fourier transform of the time derivatives in the frequency domain. We finally need to take the average in the stationary state and symmetrize the result as dictated by ${\cal T}_S$. 

With the diagrammatic approach we realize that the diagrams contributing to the final result can be arranged in tables (see Fig.~\ref{diagrams} (b) to (d)). These reproduce the results given in (\ref{Igreater})-(\ref{S3greater}). To construct these tables we proceed according to the following rules:
\begin{itemize}
\item To arrive to an expression for the cumulant of order $N$, write a table with $N$ time (frequency) intervals and corresponding superscripts $\tilde{k}$. The propagation of time will be taken from right to left.
\item Write all the possible diagrams having $N$ crosses (jumps) distributed in the different intervals, with the constraint that the maximum number of crosses in each is set by the corresponding index $\tilde{k}$. Diagrams with $n$ jumps occurring at the same time have to be included as well. These crosses are enclosed together with a box, and contribute with the jump operator $\mathcal{J}_0^{(n)}:=\partial_{\chi}^n\mathcal{J}_{\chi}|_{\chi=0}$.
\item Taking into account that jumps occurring in the same interval are indistinguishable, and that each cross can be associated to one of the possible counting fields $\chi_1,\ldots,\chi_k$ present in that interval, write the multiplicity of each diagram on the right. 
\item Write the mathematical expression corresponding to each diagram (see Fig.~\ref{diagrams} (a)) and sum the different terms evaluated at $z=i\omega$. 
\item Take the average in the stationary state, and multiply by $(-i)^N$. The resulting expression corresponds to the {\it unsymmetrized} (``greater'', $>$) {\it moment of the number of particles}. 
\item Multiply by $(i\omega_1)\ldots (i\omega_N)$. This gives the {\it unsymmetrized moment of the current distribution}.
\item Add the ``lesser'' ($<$) part, that is, the expression corresponding to negative frequency. 
\item Finally, symmetrize the result, adding all the possible switchings of frequencies. This gives the {\it symmetrized moment of the current distribution}.
The result can be simplified using (\ref{simplificationOmegas}) and (\ref{deltaIdentity}).
\end{itemize}

As mentioned above, explicit derivation gives the same result for the expressions of \textit{cumulants} of the current distribution as those derived for the moments up to $N=3$. To higher orders it is unknown for us if this property still holds or not.
Expressions for the cross correlations, e.g. $S^{(2)}_{LR}:= \langle I(t_1)I(t_2) \rangle_c$, between two (or more) stochastic processes, e.g.
$L$ and $R$, can also be derived with this technique. To this end we simply need to label each of the jumps occurring at $L$ or $R$ accordingly (see Fig.~\ref{diagrams} (c)), having two types of jump operators, ${\cal J}_L$ and ${\cal J}_R$. Also, expressions for the total current ($\alpha I_L +\beta I_R$) or accumulated current ($I_L-I_R$) can be derived using the jump operators (\ref{Jtot})-(\ref{Jaccum}) in the diagrams.

\subsection*{A.2. Equivalent form.}

We can write down an equivalent form to expressions (\ref{currentformula})-(\ref{skewnessformula}). This will allow us to obtain an analytical expression for their zero-frequency limit, which is not well defined in the form given above. To this end we make use of the projectors $P:=\vert 0 \rangle\!\rangle \langle\!\langle  \widetilde{0} \vert$ and $Q:=\mathds{1}-P$, where $P$ projects onto the subspace spanned by the stationary state $\vert 0 \rangle\!\rangle \equiv \bm{\rho}_\mathrm{S}^{stat}$ \footnote{The state $\langle\!\langle \widetilde{0}\vert$ denotes the left eigenvector of the Liouvillian. The tilde indicates that it is not the adjoint to $\vert 0 \rangle\!\rangle$, since the Liouvillian is not Hermitian.}; and we define the pseudo-inverse $R_0(z):= Q \Omega_0(z) Q$, such that $\Omega_0(z) = R_0(z) + P/z$.
Making this change in (\ref{Igreater})-(\ref{S3greater}) and symmetrizing the expression (including positive and negative frequencies) we get
\beq
 i \langle I (z) \rangle = \delta(z) \langle {\cal J}_0^{(1)} \rangle,
\eeq
\beq
 i^2 S^{(2)}(z_1,z_2)= \delta(z_1+z_2)
  \langle {\cal J}_0^{(2)} + {\cal J}_0^{(1)}R_0(z_1){\cal J}_0^{(1)} + {\cal J}_0^{(1)}R_0(z_2) {\cal J}_0^{(1)} \rangle, \nonumber\\
\eeq
\beq
  i^3 S^{(3)}(z_1,z_2,z_3) &=& \delta(z_1+z_2+z_3) \nonumber\\
  &&\times\langle
	  {\cal J}_0^{(3)}
	  +
	  {\cal J}_0^{(1)} \rb{R_0(z_1) + R_0(z_2)+R_0(z_3) }{\cal J}_0^{(2)} \nonumber \\
	  &&+
	  {\cal J}_0^{(2)}  \rb{R_0(z_{12}) + R_0(z_{23})+R_0(z_{13}) }{\cal J}_0^{(1)}
	\nonumber\\
	&&
		+ 
		{\cal J}_0^{(1)} R_0(z_1) {\cal J}_0^{(1)} \rb{R_0(z_{12}) + R_0(z_{13})} {\cal J}_0^{(1)} \nonumber \\
	&&	+ 
		{\cal J}_0^{(1)} R_0(z_2) {\cal J}_0^{(1)} \rb{R_0(z_{12}) + R_0(z_{23})} {\cal J}_0^{(1)}
	\nonumber\\
	&&
		+ 
		{\cal J}_0^{(1)} R_0(z_3) {\cal J}_0^{(1)} \rb{R_0(z_{13}) + R_0(z_{23})} {\cal J}_0^{(1)}
	\nonumber\\
	&&
	  +
	  z_1^{-1} \eww{{\cal J}_0^{(1)}} 
	  {\cal J}_0^{(1)} \left[ R_0(z_{12}) \right. \nonumber\\ 
	  &&\left. -R_0(z_2)+ R_0(z_{13}) -R_0(z_3) \right] {\cal J}_0^{(1)}
	\nonumber\\
	&&
	  +
	  z_2^{-1} \eww{{\cal J}_0^{(1)}} 
	  {\cal J}_0^{(1)} \left[ R_0(z_{12}) \right. \nonumber\\
	  && \left. -R_0(z_1)+ R_0(z_{23}) -R_0(z_3) \right] {\cal J}_0^{(1)}
	\nonumber\\
	&&
	  +
	  z_3^{-1} \eww{{\cal J}_0^{(1)}} 
	  {\cal J}_0^{(1)} \left[ R_0(z_{13}) \right. \nonumber\\
	  && \left. -R_0(z_1)+ R_0(z_{23}) -R_0(z_2) \right] {\cal J}_0^{(1)}
	\rangle.
\eeq
Now we make use of the delta function to write $z_2=-z_1$ in the noise expression and $z_3=-z_1-z_2$ in the skewness result. Performing the change of variables $z_1 \to -i \omega$, $z_2 \to i \omega$ in the noise and $z_1\to -i\omega$, $z_2\to i\omega-i\omega'$, $z_3\to i\omega'$ in the skewness, we obtain
\beq \label{IRw}
 i I_{stat} = \langle {\cal J}_0^{(1)} \rangle,
\eeq
\beq \label{S2Rw}
 i^2 S^{(2)}(\omega)=
  \langle {\cal J}_0^{(2)} + {\cal J}_0^{(1)}R_0(i\omega){\cal J}_0^{(1)} + {\cal J}_0^{(1)}R_0(-i\omega') {\cal J}_0^{(1)} \rangle,
\eeq
\beq \label{S3Rw}
  i^3 S^{(3)}(\omega,\omega') &=& 
  \langle
	  {\cal J}_0^{(3)}
	  +
	  {\cal J}_0^{(1)} \rb{R_0(-i\omega) + R_0(i\omega-i\omega')+R_0(i\omega') }{\cal J}_0^{(2)} \nonumber \\
	  &&+
	  {\cal J}_0^{(2)}  \rb{R_0(-i\omega') + R_0(i\omega)+R_0(i\omega'-i\omega) }{\cal J}_0^{(1)}
	\nonumber\\
	&&
		+ 
		{\cal J}_0^{(1)} R_0(-i\omega) {\cal J}_0^{(1)} \rb{R_0(-i\omega') + R_0(i\omega'-i\omega)} {\cal J}_0^{(1)} \nonumber \\
	&&	+ 
		{\cal J}_0^{(1)} R_0(i\omega-i\omega') {\cal J}_0^{(1)} \rb{R_0(-i\omega') + R_0(i\omega)} {\cal J}_0^{(1)}
	\nonumber\\
	&&
		+ 
		{\cal J}_0^{(1)} R_0(i\omega') {\cal J}_0^{(1)} \rb{R_0(i\omega'-i\omega) + R_0(i\omega)} {\cal J}_0^{(1)}
	\nonumber\\
	&&
	  +
	  (-i\omega)^{-1} \eww{{\cal J}_0^{(1)}} 
	  {\cal J}_0^{(1)} \left[ R_0(-i\omega') \right. \nonumber\\ 
	  &&\left. -R_0(i\omega-i\omega')+ R_0(i\omega'-i\omega) -R_0(i\omega') \right] {\cal J}_0^{(1)}
	\nonumber\\
	&&
	  +
	  (i\omega-i\omega')^{-1} \eww{{\cal J}_0^{(1)}} 
	  {\cal J}_0^{(1)} \left[ R_0(-i\omega') \right. \nonumber\\
	  && \left. -R_0(-i\omega)+ R_0(i\omega) -R_0(i\omega') \right] {\cal J}_0^{(1)}
	\nonumber\\
	&&
	  +
	  (i\omega')^{-1} \eww{{\cal J}_0^{(1)}} 
	  {\cal J}_0^{(1)} \left[ R_0(i\omega'-i\omega) \right. \nonumber\\
	  && \left. -R_0(-i\omega)+ R_0(i\omega) -R_0(i\omega-i\omega') \right] {\cal J}_0^{(1)}
	\rangle.
\eeq
The limit $\omega\to 0$ of these expressions is well defined, and they can therefore be used to check that the proper result is recovered in that limit.

\subsection*{A.3. Zero-frequency limit.}

As mentioned, expressions (\ref{IRw})-(\ref{S3Rw}) are well behaved when $\omega\to 0$. The zero-frequency noise comes straightforwardly from (\ref{S2Rw}) setting $\omega=0$. For the skewness, this limit requires nevertheless noticing that
\beq
\lim_{\omega\to 0} \left[R_0(i\omega) - R_0(-i\omega)\right] = 2i \omega \partial_\omega R_0(i\omega)\vert_{\omega=0}.
\eeq
So the zero-frequency skewness can be written as
\beq
iS^{(3)}(0,0) &=&  \eww{
	  {\cal J}_0^{(3)}
	  +
	  3 {\cal J}_0^{(1)} R_0(0){\cal J}_0^{(2)}
	  +
	  3 {\cal J}_0^{(2)} R_0(0) {\cal J}_0^{(1)} \nonumber\\
		&&+ 6 {\cal J}_0^{(1)} R_0(0) {\cal J}_0^{(1)} R_0(0) {\cal J}_0^{(1)}} \nonumber\\
		&&+   6 \eww{{\cal J}_0^{(1)}}\eww{{\cal J}_0^{(1)} \partial_\omega R_0(0) {\cal J}_0^{(1)}}.
\eeq
Now, since $\partial_\omega R_0(0) = R(0) R(0)$, we have 
\beq
iS^{(3)}(0,0) &=&  \eww{
	  {\cal J}_0^{(3)}
	  +
	  3 {\cal J}_0^{(1)} R_0(0){\cal J}_0^{(2)}
	  +
	  3 {\cal J}_0^{(2)} R_0(0) {\cal J}_0^{(1)} \nonumber\\
		&&+ 6 {\cal J}_0^{(1)} R_0(0) {\cal J}_0^{(1)} R_0(0) {\cal J}_0^{(1)}} \nonumber\\
		&&+   6 \eww{{\cal J}_0^{(1)}}\eww{{\cal J}_0^{(1)} R_0(0) R_0(0) {\cal J}_0^{(1)}}.
\eeq
Which is the zero-frequency limit found in \cite{flindtepl}.

\section{Derivation of the self-energy} \label{appKernel}

To calculate the kernel of equation (\ref{EOMchi}), we follow the
perturbative treatment by Schoeller and coworkers
\cite{Schoeller09,Leijnse08}. Let $\mathcal{L}_S$, $\mathcal{L}_R$
and $\mathcal{L}_T$ be the corresponding Liouvillians to
(\ref{Hs})-(\ref{Hv}). The last can be written in the form \beq
\mathcal{L}_T = -i \sum_{\eta,\alpha,m,\xi,p} {\cal V}_{\eta\alpha m}
G_{m}^{\xi p} J_{\eta\alpha}^{\xi p}(\chi), \eeq where $\xi = +$
$(-)$ refers to the creation (annihilation) of particles in the
leads, and $G_{m}^{\xi p}$ and $J_{\eta\alpha}^{\xi p}(\chi)$ are system
and reservoir superoperators respectively, that act on an operator
$A$ as \beq
G_{m}^{\xi p} A = \left\{ \begin{array}{lcl} \;\;\; g_m ^{\xi} A & \mbox{if} & p=+ \\
                                -A g_m ^{\xi} & \mbox{if} & p=- \end{array} \right.
\eeq

\beq
J_{\eta\alpha}^{\xi p}(\chi) A = \left\{ \begin{array}{lcl} \;\;\; j_{\eta\alpha}^{\xi}(\chi) A & \mbox{if} & p=+ \\
                                              -A j_{\eta\alpha}^{\xi}(\chi) & \mbox{if} & p=- \end{array} \right.
\eeq Where $g_{\eta\alpha} ^{\xi} $ and $j_m ^{\xi} $ are defined as
\beq
&&g_m^{+} = \sum_{aa'} \langle a \vert d_m \vert a' \rangle \vert a \rangle \langle a' \vert, \\
&&j_{\eta\alpha}^{+}(\chi) = c_{\eta \alpha}^{\dagger} e^{i
s_{\alpha}\chi_{\alpha}/2}, \eeq and $g_m^{-} =
(g_m^{+})^{\dagger}$, $j_{\eta\alpha}^{-}(\chi) = \left(
j_{\eta\alpha}^{+}(\chi) \right)^{\dagger}$. The index $s_{\alpha}=\pm 1$ is taken according to the sign convention for the current flow in lead $\alpha$.

With this notation, the self energy to order $|{\cal V}|^2$
reads \cite{Schoeller09,Leijnse08,Emary09}:
\beq \label{selfenergy}
\Sigma^{(2)} (z,\chi) \rho_S (t_0) \nonumber \\= \sum \mathrm{Tr_R} \Big\lbrace {\cal V}_{\eta\alpha m} G_{m}^{\xi p} J_{\eta \alpha}^{\xi p}(\chi) \frac{-1}{z-\mathcal{L}_S-\mathcal{L}_R} {\cal V}_{\eta' \alpha' m'} G_{m'}^{\xi' p'} J_{\eta' \alpha'}^{\xi' p'}(\chi) \rho(t_0) \Big\rbrace \nonumber \\
= \frac{1}{2\pi} \sum  \int_{-D}^{D} \Gamma_{\alpha m m'}^{\xi p p'} (\varepsilon,\chi) G_m^{\xi p} \nonumber\\ \times \frac{-p'}{z -i\xi (\varepsilon + \mu_{\alpha}) -i\lambda_a} \vert a \rangle\!\rangle \langle\!\langle \widetilde{a} \vert G_{m'}^{-\xi p'} f(-\xi p \varepsilon/kT) d\varepsilon \rho_S(t_0) \nonumber \\
= \sum -pp' \Gamma_{\alpha m m'}^{\xi p p'}(\varepsilon,\chi) G_m^{\xi p} \vert a \rangle\!\rangle \langle\!\langle \widetilde{a} \vert G_{m'}^{-\xi p'} \nonumber\\ \times \Big[ \frac{1}{2} f\left( p (\lambda_a +\xi\mu_{\alpha} -iz) \right) + \frac{ip}{2\pi} \phi\left( p (\lambda_a +\xi\mu_{\alpha} -iz) \right) \Big] \rho_S(t_0), \nonumber \\
\eeq 
where the summations run over all scripts, and $D$ is a high-energy cutoff set by the bandwidth of the Fermi leads -- larger than the rest of energy scales in the problem. 
In this expression, we have
introduced a complete set of eigenstates of the system
Liouvillian, $\mathcal{L}_S \vert a \rangle\!\rangle = i \lambda_a \vert a
\rangle\!\rangle$, and the definitions $f(x) := (e^{x/kT} +1)^{-1}$ and
\beq \label{Gammachi}
&&\Gamma_{\alpha m m'}^{\xi p p'}(\varepsilon,\chi) := \Gamma_{\alpha m m'} (\varepsilon) e^{i s_{\alpha} \xi \left(\frac{p'-p}{2}\right) \chi_{\alpha}}, \\
&&\phi(x) := \mbox{Re} \; \Psi \left( \frac{1}{2} + i \frac{x}{2\pi kT}
\right) - \mbox{ln} \; \frac{D}{2\pi kT}, \eeq being 
$ \Gamma_{\alpha m m'} (\varepsilon) \equiv \frac{2\pi}{\hbar} \sum_{\eta} {\cal V}_{\eta\alpha m} {\cal V}_{\eta
\alpha m'} \delta(\varepsilon - \varepsilon_{\eta\alpha})$ (which we take to be independent of the energy $\Gamma_{\alpha m m'} (\varepsilon)\approx \Gamma_{\alpha m m'}$), and $\Psi$ is the digamma function. The self-energy (\ref{selfenergy}) is important as it permits us to explore correctly the low bias limit ($eV\lesssim kT$) to sequential tunneling order. This self-energy is non-Markovian as the Markovian approximation has not been made up to this point. This can be made (together with the secular approximation) taking the limit $z\to 0$ of (\ref{selfenergy}), and it is what has been used throughout the text.

\end{document}